\documentclass[lettersize, journal]{IEEEtran}
\usepackage{amsmath,amsfonts}
\usepackage{algorithmic}
\usepackage{algorithm}
\usepackage{array}
\usepackage[caption=false,font=normalsize,labelfont=sf,textfont=sf]{subfig}
\usepackage{textcomp}
\usepackage{stfloats}
\usepackage{url}
\usepackage{verbatim}
\usepackage{graphicx}
\usepackage{booktabs}
\usepackage{orcidlink}
\usepackage[backend=bibtex,style=ieee,natbib=true]{biblatex} 
\begin{document}




\title{Exploring Perception-Based Techniques for Redirected Walking in VR: A Comprehensive Survey}


\author{
Bradley Coles\,\orcidlink{0009-0000-3389-3333}, 
Yahya Hmaiti\,\orcidlink{0000-0003-1052-1152}, 
and Joseph J. LaViola Jr.\,\orcidlink{0000-0003-1186-4130}%
\thanks{B. Coles is with the University of Central Florida, Orlando, Florida, USA (email: br015930@ucf.edu).}%
\thanks{Y. Hmaiti is with the University of Central Florida, Orlando, Florida, USA (email: Yohan.Hmaiti@ucf.edu).}%
\thanks{J. J. LaViola Jr. is with the University of Central Florida, Orlando, Florida, USA (email: jlaviola@ucf.edu).}%
}


\maketitle


\begin{abstract}
We present a comprehensive survey of perception-based redirected walking (RDW) techniques in virtual reality (VR), presenting a taxonomy that serves as a framework for understanding and designing RDW algorithms. RDW enables users to explore virtual environments (VEs) larger than their physical space, addressing the constraints of real walking in limited home VR setups. Our review spans 232 papers, with 165 included in the final analysis. We categorize perception-based RDW techniques based on gains, gain application, target orientation calculation, and optional general enhancements, identifying key patterns and relationships. We present data on how current work aligns within this classification system and suggest how this data can guide future work into areas that are relatively under explored. This taxonomy clarifies perception-based RDW techniques, guiding the design and application of RDW systems, and suggests future research directions to enhance VR user experience.
\end{abstract}






\section{Introduction}

\IEEEPARstart{L}{ocomotion} in Virtual Reality (VR) remains a foundational challenge for reaching the full benefits VR affords. While virtual environments (VEs) often span expansive or even infinite spaces, the real-world physical environments (PEs), especially domestic settings, are typically constrained. This mismatch between the physical and virtual spaces brought about the development of various locomotion techniques, ranging from artificial navigation methods (i.e. joystick-based movement) to techniques affording real walking~\cite{Nilsson18b}. Among these, real walking consistently provides the most compelling sense of presence, a user's sense of being there, and spatial understanding in VEs~\cite{Langbehn18a, Kim24, Skarbez, Belga_2025, hmaiti2024visual}.

Redirected Walking (RDW) was introduced as a solution to enable real walking in VEs that exceed physical boundaries, with the seminal zigzag technique introduced by Razzaque et al. \cite{Razzaque01}. Central to RDW is altering the mapping between virtual and physical locomotion, encouraging users to adjust their movements without conscious awareness. A notable early implementation consists of the Steer-to-Center (S2C) technique, which applied rotational and curvature gains to redirect users towards the center of the tracking space~\cite{Razzaque05}.

Building upon foundational RDW techniques, current RDW algorithms diversified into various approaches, including \textit{Artificial Potential Field} (APF) based avoidance techniques like APF Redirected Walking (APF-RDW)~\cite{Bachmann19}, alignment techniques supporting passive haptics~\cite{Thomas20b}, and predictive techniques leveraging predictions about future locomotion using reinforcement learning~\cite{Lee19} or graph-based optimization~\cite{Zmuda13}. These approaches primarily fall into perception-based RDW techniques manipulating users’ sense of motion through gains, contrasting with environment-based approaches that can either distort the VE to fit within the bounds of the physical environment, or take advantage of change blindness to create a paradox of multiple rooms fitting within the physical bounds \cite{Li22, Fan23a, Nilsson18a}. While environment-based techniques were extensively covered in prior surveys and perception-based techniques have received considerable attention, a gap exists in organizing and classifying sub-components of perception-based RDW techniques in a way that supports the development of future RDW techniques.

To address this gap, we propose a taxonomy for perception-based RDW techniques, focusing on their modular components and introducing a neglected classification criterion: \textit{Target Orientation Calculation}. This component defines how RDW techniques determine the physical direction users need to be guided towards to achieve successful redirection. This component is required for redirection to occur and has not been explicitly identified and categorized in prior literature, with previous surveys focusing primarily on the algorithm type rather than orientation method \cite{Li22, Nilsson18a, Fan23a, Azmandian22a}. Algorithm type refers to the method that the redirection is applied to the user's motions: \textit{Reactive} or sometimes \textit{Generalized} techniques react to the user's current state applying redirection to the user's most recent motion to direct their heading towards the calculated ideal direction \cite{Li22, Nilsson18a, Fan23a, Azmandian22a}. \textit{Predictive} techniques add a `future awareness' to this by predicting the user's future locomotion so that redirection is applied to avoid collisions on this predicted future path \cite{Li22, Nilsson18a, Fan23a, Azmandian22a}. \textit{Scripted} techniques extend this future awareness by explicitly scripting all redirection to occur within the environment using some technique (e.g. mapping) to pre-calculate the ideal redirection at any given state, requiring no calculation during operation \cite{Li22, Nilsson18a, Fan23a, Azmandian22a}. However, these categorizations do not consider how the algorithm determines the ideal orientation to redirect users towards as there can be various methods of doing so even between two algorithms of the same type. Thus, we present \textit{Target Orientation Calculation} as a new component that explicitly defines and categorizes this portion of the redirection. Through our taxonomy we provide a comprehensive framework for researchers and developers to systematically and comprehensively analyze existing methods, and also construct new RDW techniques.

We conducted an extensive analysis of 164 papers selected from an initial pool of 232 studies, we synthesized current RDW techniques and categorized them under our taxonomy. Our work addresses limitations in prior classifications by emphasizing overlooked components, including \textit{Target Orientation Calculation}, and shedding light on their impact on the effectiveness of RDW techniques. We found that there is a major gap in the current corpus under the new taxonomy we propose as there were no techniques using alignment Target Orientation Calculation and predictive Gain Application. We also found that 32.1\% of techniques use the combination of steering Target Orientation Calculation and reactive Gain Application, this makes sense since the foundational methods of RDW used these. We also found that 54.5\% of works use steering Target Orientation Calculation, and 63.2\% use reactive Gain Application. We list our contributions below:

\begin{itemize} 
    \item A novel classification system for perception-based RDW techniques, focusing on the underexplored component of Target Orientation Calculation, addressing limitations in prior taxonomies. 
    \item A modular component-wise framework for constructing RDW techniques and contextualizing/organizing existing ones, offering a practical tool for researchers or developers.
    \item An in-depth discussion of perception-based RDW techniques, contextualized within the taxonomy, to highlight the state of the art in the RDW field. 
    \item An identification of key avenues for future research, informed by our analysis of RDW literature, taxonomy, and gaps in current methodologies. 
\end{itemize}

\section{Related Work}

Prior work on RDW includes existing taxonomies and surveys. Many of these taxonomies define algorithms as being \textit{reactive}, \textit{predictive}, or \textit{scripted}~\cite{Azmandian22a, Fan23a, Nilsson18a}. \textit{\textbf{Reactive}} or \textit{\textbf{generalized}} algorithms are described as those that require no prediction of user locomotion, or pre-scripting, and instead react to the user's current state within the VE~\cite{Azmandian22a, Fan23a, Nilsson18a}. \textit{\textbf{Predictive}} or \textit{\textbf{dynamic planning}} algorithms make a prediction about user locomotion and use this predicted path to determine the ideal redirection to apply to the user to redirect this future motion to an ideal state~\cite{Azmandian22a, Fan23a, Nilsson18a}. \textit{\textbf{Scripted}} or \textit{\textbf{static planning}} algorithms require a pre-scripted pathway or environmental mapping to determine redirections to apply to users at a given state and do not work when users deviate from this scripted state~\cite{Azmandian22a, Fan23a, Nilsson18a}. Li et al. \cite{Li22} expanded on these categories by introducing reinforcement learning as a distinct class.

Steinicke et al. \cite{Steinicke08b} proposed a taxonomy of RDW based on their defined \textit{locomotion triple $(s,u,w)$}: defined by the strafe vector $s$, the up vector $u$, and the direction of walk vector $w$, while $s$ and $w$ are always orthogonal $u$ is not limited to being orthogonal to $s$ and $w$. This taxonomy is very different than our proposed taxonomy and we include a bulk more work since 2008. While the concept of the locomotion triple remains relevant to understanding gains, we believe our component-wise framework provides a more complete and useful taxonomy for perception-based RDW.

Nilsson et al. \cite{Nilsson18b} provide a broad survey of natural walking techniques, including a classification of VR locomotion methods beyond RDW, categorizing RDW into two main approaches: perception manipulation and virtual space manipulation. Similarly, Nilsson et al. \cite{Nilsson18a} uses this classification of perception manipulation and space manipulation techniques and gives an overview of work from the original RDW from Razzaque et al. \cite{Razzaque01} to 2018. This taxonomy for RDW considers only the algorithm type, which we have classified as the \textit{Gain Application} component of our framework. This taxonomy for RDW is prevalent in literature to this day and we used this as the starting point for performing our survey and deriving our taxonomy.

Li et al. \cite{Li22} also adopt the perception- vs. environment-based manipulation distinction for RDW maneuvers, emphasizing gain usage and introducing enhancements such as Reinforcement Learning (RL), multi-user support, and jumping.
They provide a RDW taxonomy encompassing environment- and perception-based techniques. 
Their classification of perception-based techniques aligns with our taxonomy, which initially categorized algorithms as reactive, predictive, or scripted, reflecting other prior taxonomies~\cite{Nilsson18a}.
However, we focus in-depth on perception-based techniques, offering a more comprehensive component-wise framework as taxonomy of these algorithms.
While they focus solely on gains and algorithm type, we also include the explicitly defined \textit{Target Orientation Calculation type} component, separating how the technique determines user redirection from the method used to apply gains: \textit{reactive}, \textit{predictive}, and \textit{scripted}.
We, therefore, enhance the possible understanding and analysis of RDW research and techniques by explicitly separating and identifying this component. We provide a new component and include other similar components to other taxonomies, such as Li et al. \cite{Li22}, to ensure completeness of taxonomy and analysis.
We believe their taxonomy is limited in its ability to guide new RDW research.
Their classification does not clarify how to explore new approaches, as algorithms can be both be both learning-based and reactive, yet are treated as distinct categories.
As such, their taxonomy does not effectively guide the creation of new RDW techniques, limiting its usefulness for further research. In contrast, our framework provides a more comprehensive and practical structure for developing perception-based RDW algorithms.
Another limitation of their survey is the lack of transparency in the collection and review of papers. We address this by detailing our collection, review, and taxonomy creation processes, enabling future work to extend our survey.

Fan et al. \cite{Fan23a} provide a survey focused on the distinction between traditional methods such as S2C and novel methods such as APF-RDW and also provide an overview of what experimental metrics are used in the literature to evaluate RDW methods. Their survey is similar in content to Li et al. \cite{Li22}, however, they provide a different taxonomy. Similarly, we believe that our taxonomy goes a step even further and provides a more complete framework for perception-based RDW. Though, as with these other surveys they consider the significance of environment-based RDW techniques to the field, whereas we limit our work to only perception-based RDW. We address similar limitations of this survey on the clarity of the survey process and use for construction of new and implementation of existing RDW algorithms.

The survey from Gemert et al. \cite{Gemert24} focuses specifically on VR locomotion, including but not limited to RDW, and its relationship to simulator sickness. While they provide a great analysis of this, they focus solely on the simulator sickness of these various locomotion techniques and focus on a broad spectrum of techniques (i.e. teleportation, real walking, joystick, and RDW), limiting the ability to perform deeper analysis on specific techniques. Instead, we focus narrowly on perception-based RDW techniques to outline a framework for the component-wise construction of perception-based RDW techniques.

\section{Methodology}
After iterating through prior work about RDW and related fields, we developed a taxonomy to classify perception manipulation RDW techniques. This taxonomy is intended to guide the classification of current and future RDW techniques, identify gaps for further exploration, and assist in selecting appropriate methods for specific applications. Below in this section we detail our systematic approach to compiling relevant works, defining our inclusion and exclusion criteria, and the process for defining our taxonomy. Systematic reviews employ structured methodologies to reduce bias and enhance the reliability of the findings, and they are crucial in the validation of existing practices, discovering new approaches, and guiding future research efforts~\cite{chandler2019cochrane, munn2018systematic}.
We followed the steps suggested by the PRISMA guidelines to conduct a systematic review~\cite{Pagen71}, and note that the filtering, review of papers, and taxonomy creation were performed by the primary author and reviewed by the other authors.

To build a comprehensive archive of relevant works, we conducted a systematic comprehensive search across several online academic digital libraries and repositories, including \textit{ACM Digital Library}, \textit{IEEE Xplore}, and \textit{Springer}.
To optimize our search for most relevant results, we used targeted keywords including: "Redirected Walking", "RDW", "Virtual Reality", "VR", "Mixed Reality", "MR", "Augmented Reality", "AR", "Extended Reality" and "XR". We formulated the following search query: \textbf{("Redirected Walking" OR "RDW") AND ("Virtual Reality" OR "Mixed Reality" OR "Augmented Reality" OR "Extended Reality" OR "VR" OR "MR" OR "AR" OR "XR")}. The search was case-insensitive to ensure inclusion of all relevant results. This yielded a total of 1,824 papers, 519 papers from \textit{ACM Digital Library}, 566 papers from \textit{IEEE Xplore}, and 739 papers from \textit{Springer}, providing a broad and diverse set of articles for review.

After collecting all the papers, we systematically filtered out those irrelevant to our taxonomy. To ensure relevance and quality in this process, we initially established predefined clear exclusion and inclusion criteria for the collected papers~\cite{Pagen71}. We defined two inclusion criteria (IN): "The paper presents a study on \textit{Redirected Walking}" (IN-1) and "The paper presents a study on \textit{Locomotion Gain Detection}" (IN-2). These exclusion (EX) criteria were refined as we progressed through multiple iterations. The exclusion criteria are listed below:

\begin{description}
\item [EX-1] \textit{Paper is not in English or inaccessible.}
\item [EX-2] \textit{Paper is a poster, short paper, or lacks peer review.}
\item [EX-3] \textit{Paper is off-topic: Paper is unrelated to VR/AR/XR, does not address locomotion, or does not focus on human-centered interaction (e.g. robot locomotion)}.
\item [EX-4] \textit{Paper does not provide a contribution to perception-based RDW or just mentions it but is focused on another topi, such as reviews or overviews without new experimental results or theoretical improvements. Environment-based RDW techniques were omitted from this survey in order to focus on perception-based RDW. Contributions were defined by providing new experimental results for new algorithms, enhancements to existing algorithms, results contradictory to prior findings, or surveys producing novel insights.}
\end{description} 

Following this iteration and based on guidelines presented by levac et al. \cite{levac2010scoping}, we evaluated every title and abstract against our criteria and only kept papers deemed relevant. We assessed the study quality of each kept paper through a more refined assessment, such that we considered a paper as high-impact if it presented a new RDW technique, algorithm, user study, or theoretical advancements that improve the understanding or performance of RDW across the mixed reality spectrum. Papers merely mentioning RDW without proving new insights or improvements were omitted under [EX-4]. In cases of uncertainty, we examined the full paper to make an informed decision about its inclusion. Papers meeting the exclusion criteria were excluded and all the rest retained. The papers excluded, yet with slight relevance to our topic were noted for potential future reference. After performing this iteration, our final relevant papers corpus consisted of 232 papers.

After filtering the initial set of papers, we reviewed each individually, noting the title, link, inclusion status, main contributions, and relevant taxonomy classifications. We further refined this list of papers by assessing the impact of each paper, focusing on high-impact works published in reputable venues and widely cited within the RDW community. We reviewed the full text of these papers against the exclusion criteria to determine relevance, discarding any that met an exclusion criterion. The list of papers following this totaled 162. We then used the following inclusion criteria to include further works that were not found within the scope of the search on the archives:
\begin{description}
\item [IN-3] \textit{Paper was highly cited by papers in the corpus and provided important RDW definitions or findings}.
\item [IN-4] \textit{Paper was recommended to authors and after review met no exclusion criteria}.
\end{description}

After including papers following these criteria, 3 new papers were added to the corpus resulting in a total of 165 papers. Our goal was to survey the current work on RDW, present a taxonomy and summary of the field, and identify potential avenues for future research. We ensured high-quality selection by avoiding low-impact papers and eliminating redundancies while refining the taxonomy. The exclusion of low-impact papers affords focusing on works that shaped the field of RDW and ensuring that the taxonomy is built upon well validated and widely accepted and used research. The inclusion of low-impact works might dilute the quality of the taxonomy and introduce unverified concepts that were not peer-reviewed rigorously.

During the initial filtering of papers, we started with an initial taxonomy of algorithm type as \textit{Reactive}, \textit{Predictive}, or \textit{Scripted} as is commonly seen in the RDW literature. 
We categorized each included paper and refined the taxonomy during the review process. Notably, some papers did not fit a specific algorithm type, as they used multiple algorithms or proposed improvements applicable across various types.
We analyzed notes on paper contributions and the structure of RDW algorithms, developing a revised taxonomy for reclassifying the papers.
During this revision, we found that RDW algorithms contain multiple components in order to function, namely: \textit{Gains}, \textit{Target Orientation Calculation}, and \textit{Gain Application}. We also noted that some contributions belonged outside the scope of these required components and could be applied to a variety of techniques without regard to component types. From this observation, we derived the \textit{Enhancements} component. We then formed a component-wise framework for the construction of perception-based RDW techniques as the basis for our taxonomy.

\section{Taxonomy}

\begin{figure*}[t]
  \centering
\includegraphics[width=.7\textwidth]{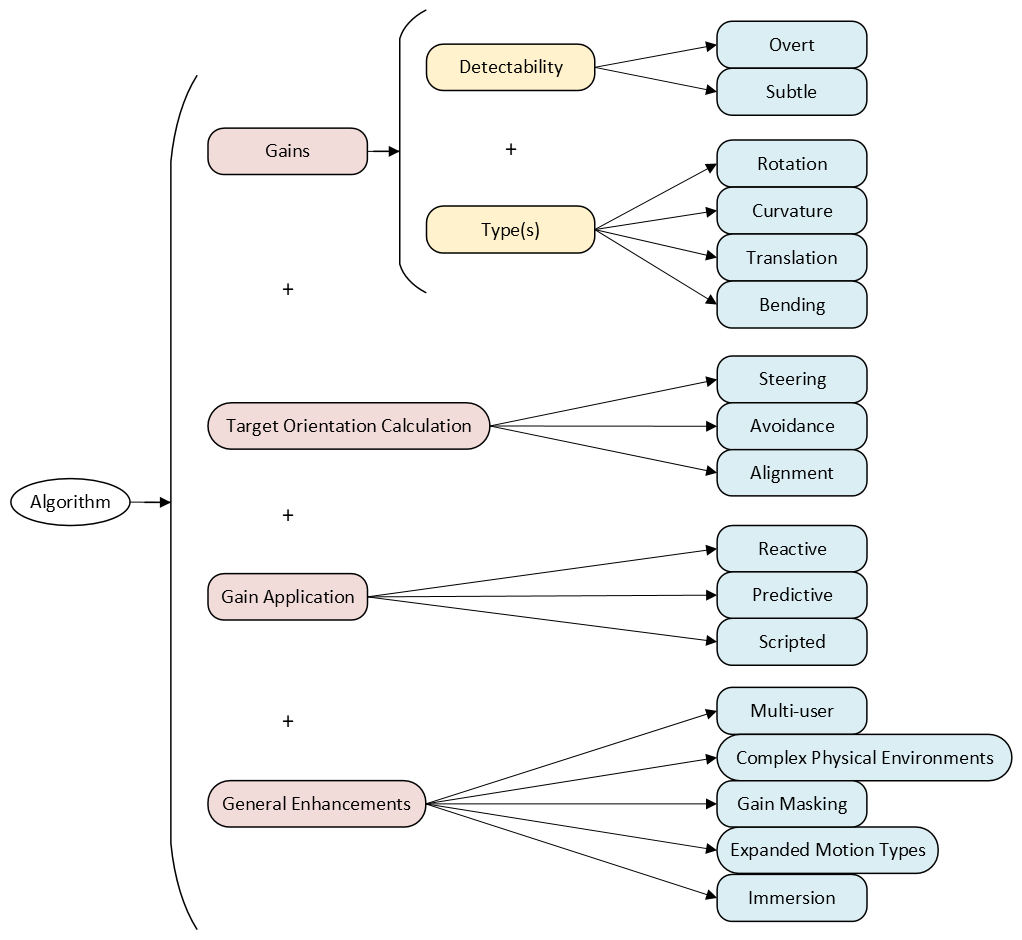}
  \caption{Taxonomy of Redirected Walking Algorithms – RDW algorithms consist of three core components: (1) locomotion gains, (2) methods for applying those gains, and (3) orientation calculation. Gains vary by detectability (overt/subtle) and type (rotation, translation, curvature, bending), with types combinable. Gain application methods are categorized as reactive, predictive, or scripted. Orientation strategies include steering, avoidance, and alignment. Optional enhancements address multi-user scenarios, complex spaces, gain masking, motion diversity, and immersion.}
  \label{fig:taxonomy}
  \vspace{-2mm}
\end{figure*}

We present a taxonomy of perception-based RDW algorithms defining algorithms as containing four components: gains, gain application method, target orientation calculation method, and general enhancements (see Fig.~\ref{fig:taxonomy}). For a RDW algorithm to effectively function, it must first decide how to apply gains to user locomotion—whether reactively based on the user's current state, predictively based on anticipated movement, or through scripted paths using environmental mapping. Afterward, the RDW algorithm needs to determine the direction to redirect users. These redirection methods include \textit{steering methods} that guide users toward specific targets, \textit{avoidance methods} steer users away from obstacles, and \textit{alignment methods} maintain environmental alignment. Lastly, the algorithm applies gains or modifications to user locomotion to direct the user toward the calculated heading. Historically, RDW was categorized either as reactive, scripted, or predictive \cite{Azmandian22a, Fan23a, Nilsson18a}. Prior surveys emphasized this categorization, and while we include it in our taxonomy as a component of perception-based RDW algorithms, our primary contribution is a new focus on categorizing target orientation.

For view-manipulation-based redirection to function effectively, the algorithm must determine the target orientation for the user in the physical environment (PE) or in other words where the user's physical movement should be directed to avoid collisions and extend the distance they can walk. While all RDW algorithms require this, previous taxonomies have not considered it a distinct category. Initially, when all RDW techniques were steering-based, this distinction wasn’t necessary. However, with the development of multiple methods for calculating orientation, focusing on this aspect as a category brings about a new perspective for analysis. This highlights potential opportunities for future work and provide alternative approaches for choosing adequate RDW methods in dissimilar applications. In order to then direct users towards this heading the perception-based RDW technique needs to apply gains or modifications to the user's movement. Typically RDW algorithms limit the amount of manipulation or gain applied to movement in order to be imperceptible and maintain the illusion of the user walking in the direction they intended, as such algorithms need to determine how much gain to apply to direct users as closely as possible to the Target orientation.

Our taxonomy also outlines enhancements applicable to any perception-based RDW algorithms, including: (1) Adapting techniques for multiple users. (2) Accommodating irregularly shaped spaces. (3) Using distractors to mask gains and surpass standard detection thresholds. (4) Expanding RDW to support locomotion beyond 2D walking. (5) Immersion improvements, defined as passive haptics support and improvements to seamlessly embed the RDW technique into the application. The following sections examine prior perception-based RDW research through the lens of our taxonomy, covering gains, the novel \textit{Target Orientation Calculation} component, gain application types, and general enhancements.

\section{Gains}
\label{sec:gains}

Gains are applied to user locomotion so that movement is not mapped one to one from the PE to the VE to achieve user redirection~\cite{Razzaque05}. Research extensively explored gains values for such applications, focusing on user discomfort and detection of locomotion manipulation (see citations in Table~\ref{tab:dts}). RDW techniques can apply gains \textit{overtly}~(see~Sec.~\ref{sec:overtGains}) or \textit{subtly}~(see Sec.~\ref{sec:subtleGains}) by leveraging information about user detection of gains.

The four types of gains typically used for RDW are \textit{rotation}, \textit{curvature}, \textit{translation}, and \textit{bending}. RDW techniques can use one or more types of gains to achieve redirection. \textit{Rotation gains} are applied to stationary user rotation, \textit{Curvature gains} are used to turn straight paths into curved ones, and \textit{Translation gains} alter the displacement of user movement, breaking the 1:1 mapping so that a user may travel further in the VE than in the PE, or vice versa~\cite{Steinicke08a}. Translation gains are used in some techniques to minimize tracking space required to navigate larger VEs~\cite{Zhang13a}. \textit{Bending gains} are a more recent concept used to further bend already curved paths, unlike curvature gains that curve straight paths~\cite{Langbehn17}. Research also investigated extending gains beyond walking movements by applying them to jumps~\cite{Hayashi19, Jung19}, in addition to slopes in the PE to adjust the experienced slope to match a slope of a different angle in the VE~\cite{Hu19}.

\subsection{Subtle Gains}
\label{sec:subtleGains}
Most RDW techniques focus on subtle gains, to redirect the user imperceptibly, so users can walk infinitely without encountering obstacles~\cite{Razzaque05}. Detectable gains degrade user experience~\cite{Kjenstad23}, or correlate with task performance in certain types of tasks~\cite{Rothacher18}. Since detectable gains can be disorienting and unnatural, extensive research focused on determining detection thresholds (DTs) for these gains. Table~\ref{tab:dts} shows notable gain detection thresholds from literature per gain type. DTs represent values within which gains cannot be reliably detected. These are typically measured using a \textit{two-alternative forced choice} (2AFC) task, where participants state the direction in which gains are applied~\cite{Steinicke08a}. For example, for curvature gains, participants are asked if the movement curves left or right. DTs represent the range from a 25\% to 75\% chance of making the correct choice~\cite{Steinicke08a}.

Steinicke et al. \cite{Steinicke08a} proposed some of the first DTs for curvature, rotation, and translation gains as shown in Table~\ref{tab:dts}. 
This was followed up in subsequent works that proposed new values for DTs~\cite{Steinicke09, Steinicke10}. Later, Langbehn et al. \cite{Langbehn17} proposed bending gains and established DTs for these gains. 
Zhang et al. \cite{Zhang18} tested thresholds for gains in 360$^{\circ}$ video teleoperation, demonstrating that scaling motion is feasible in remote teleoperation. Kim et al. \cite{Kim21} introduced the concept of relative translation gains, applying independent translation gains to both $x-$ and $y-$ axes, and defined a DT based on the ratio of these relative gains. 

Research has defined DTs for gains extending beyond two dimensional walking movement including curvature, rotation, and translation gains for jumping \cite{Hayashi19}, curvature gains for slopes \cite{Hu19, Zhang22}, and pitch and roll gains for three dimensional movement \cite{Yamamoto18, Bolte10}.
You et al. \cite{You22} introduced strafing gains, which adjust position while maintaining user orientation, unlike curvature gains. These strafing gains are more akin to translation gains, translating user's position left or right along a diagonal line without curving the path or changing orientation. Mayor et al. \cite{Mayor22} proposed deviation gains, converting rotations into translations. However, both strafing and deviation gains have seen limited adoption and testing in RDW techniques. Xu et al. \cite{Xu24c} observed that during real locomotion, oscillating bi-directional head movements occur, with rotation gains applied to each instance. The typical application of these gains can induce discomfort, so they propose \textit{Bidirectional Rotation Gain Difference} (BiRD) to leverage this aspect of locomotion. Subsequent research investigated these thresholds and identified factors influencing gain detection, to increase the magnitude of gains that can be applied imperceptibly. 

\begin{table*}
  \caption{Notable Detection Thresholds in literature}
  \label{tab:dts}
  \small 
  \centering
  \resizebox{0.9\textwidth}{!}{
  \begin{tabular}{l|ccc}
    \toprule
    Gain Type & Comment & Thresholds & Source \\
    \midrule
    Curvature & --& $r > 16$m & Steinicke et al. (2008)~\cite{Steinicke08a}\\
    & 2m path normal motion before redirection & $r > 24$m & Steinicke et al. (2008)~\cite{Steinicke08a}\\
    & -- & $r > 22$m & Steinicke et al. (2010)~\cite{Steinicke10}\\
    & Constant Stimuli & $r > 6.41$m & Grechkin et al. (2016)~\cite{Grechkin16}\\
    & maximum likelihood & $r > 11.61$m & Grechkin et al (2016).~\cite{Grechkin16}\\
    & threshold for men & $r > 10.7$m & Nguyen et al. (2018)~\cite{Nguyen18b}\\
    & threshold for women & $r > 8.63$m & Nguyen et al. (2018)~\cite{Nguyen18b}\\
    & tested left and right curves separately & $r_{right} > 10.20$m, $r_{left} > 7.81$m & Li et al. (2021)~\cite{Li21b}\\
    & post-order path & $r_{right} > 12.821$m, $r_{left} > 10.309$m & Li et al. (2021)~\cite{Li21b}\\
    & left side-step & $r_{right} > 6.02$m, $r_{left} > 13.19$m left & \citeauthor{Cho21} (2021)~\cite{Cho21}\\
    & right side-step & $r_{right} > 9.92$m, $r_{left} > 4.65$m left & \citeauthor{Cho21} (2021)~\cite{Cho21}\\
    & similar sensory sensitivity to average & $gC = 0.048$ & Matsumoto and Narumi (2022)~\cite{Matsumoto22}\\
    & higher sensory sensitivity than average & $gC = 0.037$ & Matsumoto and Narumi (2022)~\cite{Matsumoto22}\\
    & similar sensory avoidance to average & $gC = 0.048$ & Matsumoto and Narumi (2022)~\cite{Matsumoto22}\\
    & much higher sensory avoidance than average & $gC = 0.033$ & Matsumoto and Narumi (2022)~\cite{Matsumoto22}\\
    & control (no device stimuli) & $gC = -0.055$ to $0.054$ & Hwang et al. (2023)~\cite{Hwang23a}\\
    & Noisy Galvanic Vestibular Stimulation & $gC = -0.055$ to $0.061$ & Hwang et al. (2023)~\cite{Hwang23a}\\
    & Directional Galvanic Vestibular Stimulation & $gC = -0.068$ to $0.7$ & Hwang et al. (2023)~\cite{Hwang23a}\\
    & Bone-conduction Vibration & $gC = -0.061$ to $0.07$ & Hwang et al. (2023)~\cite{Hwang23a}\\
    & Caloric Vestibular Stimulation & $gC = -0.061$ to $0.068$ & Hwang et al. (2023)~\cite{Hwang23a}\\
    Rotation & -- & 0.59--1.10 & Steinicke et al. (2008)~\cite{Steinicke08a}\\
    & Subsequent rotations & 0.76--1.19 & Steinicke et al. (2008)~\cite{Steinicke08a}\\
    & -- & 0.64--1.24 & Steinicke et al. (2010)~\cite{Steinicke10}\\
    & 360-degree video teleoperation & 0.892--1.054 & Zhang et al. (2018)~\cite{Zhang18}\\
    & FOV 40 all participants w/out distractor & 0.68--1.22 & Williams and Peck (2019)~\cite{Williams19b}\\
    & FOV 110 all participants w/out distractor & 0.67--1.44 & Williams and Peck (2019)~\cite{Williams19b}\\
    & FOV 110 women only w/out distractor & 0.65--1.32 & Williams and Peck (2019)~\cite{Williams19b}\\
    & FOV 110 men only w/out distractor & 0.70--1.56 & Williams and Peck (2019)~\cite{Williams19b}\\
    & FOV 40 all participants w/ distractor & 0.68--1.25 & Williams and Peck (2019)~\cite{Williams19b}\\
    & FOV 110 all participants w/ distractor & 0.57--1.57 & Williams and Peck (2019)~\cite{Williams19b}\\
    & FOV 110 women only w/ distractor & 0.37--1.48 & Williams and Peck (2019)~\cite{Williams19b}\\
    & FOV 110 men only w/ distractor & 0.72--1.62 & Williams and Peck (2019)~\cite{Williams19b}\\
    & Jump rotation & 0.50--1.44 & \citeauthor{Hayashi19} (2019)~\cite{Hayashi19}\\
    & BiRD - oscillating head rotations & 0.84--1.24 & \citeauthor{Xu24c} (2024)~\cite{Xu24c}\\
    Translation & -- & 0.78--1.22 & Steinicke et al. (2008)~\cite{Steinicke08a}\\
    & -- & 0.86--1.26 & Steinicke et al. (2010)~\cite{Steinicke10}\\
    & 360-degree video teleoperation & 0.942--1.097 & Zhang et al. (2018)~\cite{Zhang18}\\
    & Jump distance translation & 0.68--1.44 & \citeauthor{Hayashi19} (2019)~\cite{Hayashi19}\\
    & Jump height translation & 0.09--2.16 & \citeauthor{Hayashi19} (2019)~\cite{Hayashi19}\\
    & Backward step & 0.84--1.33 & \citeauthor{Cho21} (2021)~\cite{Cho21}\\
    & DT in small empty room & 0.73--1.10 & Kim et al. (2023)~\cite{Kim23}\\
    & DT in large empty room & 0.91--1.22 & Kim et al. (2023)~\cite{Kim23}\\
    Bending & radius of real curve: $r_{real} = 1.25$m & 3.25 & Langbehn et al. (2017)~\cite{Langbehn17}\\
    & radius of real curve: $r_{real} = 2.5$m & 4.35 & Langbehn et al. (2017)~\cite{Langbehn17}\\
    Slope Curvature & real = 2\% & 1.56--5.25 & Hu et al. (2019)~\cite{Hu19}\\
    & real slope = 4\% & 1.98--4.28 & Hu et al. (2019)~\cite{Hu19}\\
    & real slope = -2\% & 1.00--7.36 & Hu et al. (2019)~\cite{Hu19}\\
    & real slope = -4\% & 1.22--3.44 & Hu et al. (2019)~\cite{Hu19}\\
    Strafing & -- & 5.57$^{\circ}$ right diagonal, 4.68$^{\circ}$ left diagonal & \citeauthor{You22} (2022)~\cite{You22}\\
    Deviation & -- & 1.74--5.60 rad/m & \citeauthor{Mayor22} (2022)~\cite{Mayor22}\\
    \bottomrule
  \end{tabular}
  }
\end{table*}

\subsubsection{Detection Factors}
\label{sec:gain_factors}
Current research on gain detection also examines factors influencing gain detection. Walking speed, shown to influence gains detection, has been controlled in many subsequent studies testing DTs~\cite{Neth12}. Interestingly, Nguyen et al. \cite{Nguyen17} found that translation gains affect walking speed, with users walking faster at gains compressing movement and slower at gains increasing movement. Other research showed rotation speed influencing rotational gains detection~\cite{Brument21}. Matsumoto et al. \cite{Matsumoto22} found that sensory sensitivity influences DTs with individuals with higher sensitivity being more likely to detect curvature gains.

Williams et al. \cite{Williams19b} investigated the effects of FOV, gender, and distractors on gain detection and found that a 110 FOV produced a wider acceptable range of gains regardless of other factors, that men and women showed significantly different thresholds for gains at 110 FOV, and that men showed less of an effect of distractors on gain detection. Further studies support the gender differences with DTs~\cite{Rothacher18}. Nguyen et al. \cite{Nguyen18b, Nguyen20a} found an effect of gender on curvature gain DTs and found a significant effect of sense of agency on curvature DTs with higher agency corresponding to higher chance of detecting redirection. Research has found that room size has a significant effect on relative translation gain detection when rooms are empty of distractors but room size does not have a significant effect with distractors present~\cite{Kim22, Kim23}. Other research has failed to find a correlation between environment appearance and gains, suggesting gains can be reliably used in both natural and urban environments~\cite{Mostajeran24}.

Grechkin et al. showed that simultaneously applied translation and curvature gains does not affect DTs and that gradually increasing curvature rate of gains allows for a larger curvature gain to be applied reducing detection~\cite{Grechkin16}. Brument et al. \cite{Brument21} showed that combining translation and rotational gains does not influence the detection of rotational gains. Li et al.~\cite{Li21b} found that sensitivity to curvature gains increases when walking along a preorder curved path. Though most work was focused on gains applied to VR using HMDs, Freitag et al. \cite{Freitag16} found that translations tested showed no significant effect to simulator sickness, presence, or cognitive load when used in a CAVE setup. Nguyen et al. \cite{Nguyen20b} found that cognitive load had a significant effect on DTs suggesting that when users are engaged in tasks higher gains than traditional DTs may be possible. Some research has shown that discomfort from curvature and bending gains decreases as gains gradually change from lower values to higher values dynamically~\cite{Sakono21}. Beyond improved comfort some research has shown that rate of change of gains is correlated to reduced sensitivity to gains and less detection~\cite{Congdon19}, while other research has shown no difference with detection~\cite{Zhang13b}. Research has also shown that users cannot reliably detect the change in gains~\cite{Lee23b, Zhang14}. Research has shown that DTs are correlated with exposure to VR and prior exposure to locomotion gains~\cite{Bolling19, Lee24e}. Research into whether avatars or virtual feet can influence detectability of gains has shown that there is no significant effect~\cite{Kruse18, Reimer20}.

Research into electrical and non-electrical stimulation to create vestibular noise in order to mask gains has shown a potential to influence DTs~\cite{Hwang23a, Hwang23b, Matsumoto21}, while other research suggests some forms of stimulation such as trans-cranial direct-current stimulation or electric muscle stimulation does not affect detection but does affect sickness~\cite{Auda19, Langbehn19}. Hwang et al. \cite{Hwang23a} found that non-electrical vestibular stimulation could increase DTs by around 19.27\% for users sensitive to electrical stimulation but that it is less safe by introducing gait instability. Kruse et al. \cite{Kruse21} found that users were more likely to detect that redirection has occurred when in an environment with co-located users due to the non-visual cues from other users in the environment such as sound. Haptics with a wall have been shown to mask detection of gains and expand DTs for both curvature and bending gains~\cite{Lee24b, Matsumoto16, Matsumoto17, Nakamura19}. Research has shown that perceptual illusions including change blindness during blinks and saccades can mask gains~\cite{Alsaeedi21, Bolte15, Joshi20, Keyvanara18, Keyvanara19, Langbehn18b, Nguyen18c, Steinicke13, Sun18}. Auditory stimuli to mask or create gains has also been tested, some research shows audio has little to no effect on rotation gain DTs when visual cues are present~\cite{Gao20, Junker21}, other research has shown that auditory cues can broaden DTs when applied or be the sole cause for user redirection when visual stimuli are missing or noisy, or when users are directed to follow sounds~\cite{Gao20, Gerritse24, Lee24a, Nogalski16, Rewkowski19, Serafin13, Weller22}. Lee et al. \cite{Lee24a} also found that olfactory stimuli could be effective for redirecting users with gains. Cho et al. \cite{Cho21} expands gains to non-forward steps determining DTs for specifically non-forward motion, finding that sidestep redirection can be redirected more towards the front of the user and backsteps show wider DTs for translation than sidesteps.

\subsection{Overt Gains}
\label{sec:overtGains}
Overt gains can be useful when subtle gains fail to prevent an imminent collision. In such cases, a reset technique is employed, prompting the user to stop and turn in place while rotation gains are applied to reorient them within the PE to a more favorable heading.~\cite{Williams07}. Simple resets like 2:1 turn perform a simple mapping of doubling user's rotation in the PE and having users rotate 360$^{\circ}$ in the VE resulting in a 180$^{\circ}$ rotation in the PE. Many RDW techniques propose custom reset techniques to operate alongside them often leveraging the same enhancements as the technique. There has also been research specifically improving reset techniques. For collaborative tasks Min et al. \cite{Min20} introduced a recovery technique using overt gains, similar to reset techniques, that preserves relative locations of two users in both the VE and PE. This allows for a handshake to occur in both VE and PE, while enabling RDW for exploring a VE larger than the PE. Zhang et al. \cite{Zhang23a} propose an out-of-place reset, a reset technique that adjusts both position and orientation, useful for narrow spaces where an in-place rotation only reset can cause frequent resets to be triggered. Lee et al. \cite{Lee24c} propose Multi-Agent Reinforcement Resetter (MARR), which uses reinforcement learning (RL) to determine reset direction showing improved performance and fewer resets in complex environments compared to more basic reset techniques as proposed by Williams et al. \cite{Williams07}, and supporting multiple users.

Thomas et al. proposed three reset methods that are commonly used as baseline reset techniques in recent research \cite{Thomas19}. These techniques show that improved reset techniques also have a major factor in the effectiveness of RDW. These techniques showed that basic reset techniques, while effective, can be improved and that subtle redirection is not the only area of concern for developing effective RDW. Li et al. \cite{Li24} propose a reset technique based on the energy potential of the environment and predicted locomotion of users. This technique uses avoidance based heading calculation and locomotion prediction for application, this illustrates how our taxonomy can be used to categorize both overt reset techniques and subtle reorientation techniques. The effectiveness of this technique also shows that there is room for expansion of reset techniques within areas of under-saturation of work based on our taxonomy of RDW techniques. Since the effectiveness of subtle RDW alone is not enough to completely prevent collisions in all cases, effective RDW requires overt reset techniques in combination with subtle redirection techniques. The important factor for these methods is not the magnitude of applied gains, as the goal is not to mask the use of these gains, but rather the way that they can be improved when used alongside subtle redirection to reduce the frequency of triggering these overt resets. These techniques also need a target heading to reorient towards, and can also be made to be reactive of the current situation or to utilize predictions of the future locomotion. It is also possible to have scripted resets as seen in Kwon's environment tiling via resets \cite{Kwon22}.

Beyond reset techniques, research shows that users may accept overt gains exceeding detection thresholds as long as sense of presence is maintained. Schmitz et al.~\cite{Schmitz18} studied rotation gain thresholds that disrupt user-reported presence and found users would accept gains that are beyond the detectable threshold for gains. Selzer et al. \cite{Selzer22} found that users would accept some translation gains above the detectable threshold. Rietzler et al. \cite{Rietzler18} showed that users would accept gains above the detectable threshold, and proposed using RDW with overt gains as a distinct interaction and locomotion technique from real walking. This suggests that for some experiences intentionally detectable gains can actually be better than without. Overall, most techniques rely on subtle gains and manipulations until they are insufficient to prevent a collision, at which point a reset technique with overt gains is initiated.

\section{Target Orientation Calculation}
\label{sec:heading_calc}

\begin{figure*}[t]
  \centering
  \includegraphics[width=.9\textwidth]{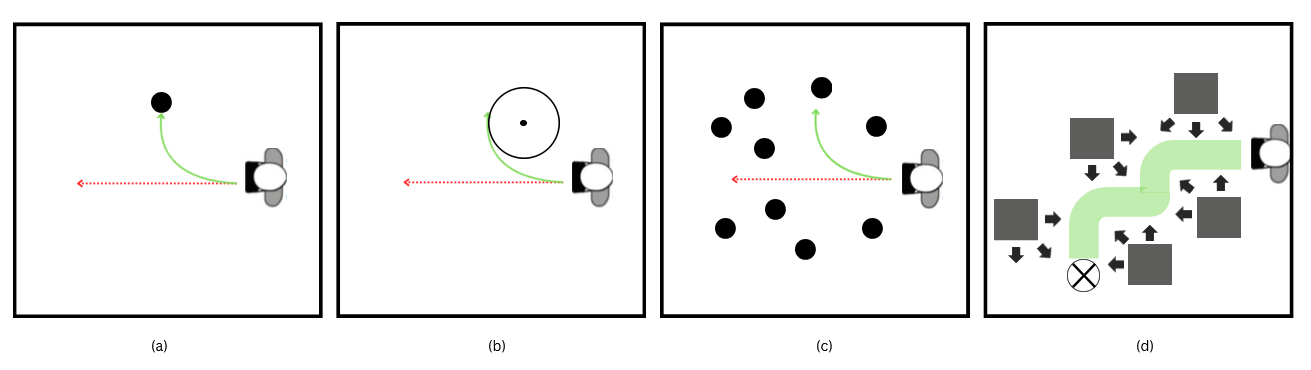}
  \vspace{-3mm}
  \caption{Steering and Avoidance Target Orientation Calculation Method Examples: a) steering to a single target. b) steering to a path. c) steering to multiple targets. d) avoidance with a path avoiding multiple obstacles.}
  \label{fig:steer-and-apf}
\end{figure*}

\begin{figure*}[t]
  \centering
  \includegraphics[width=.8\textwidth]{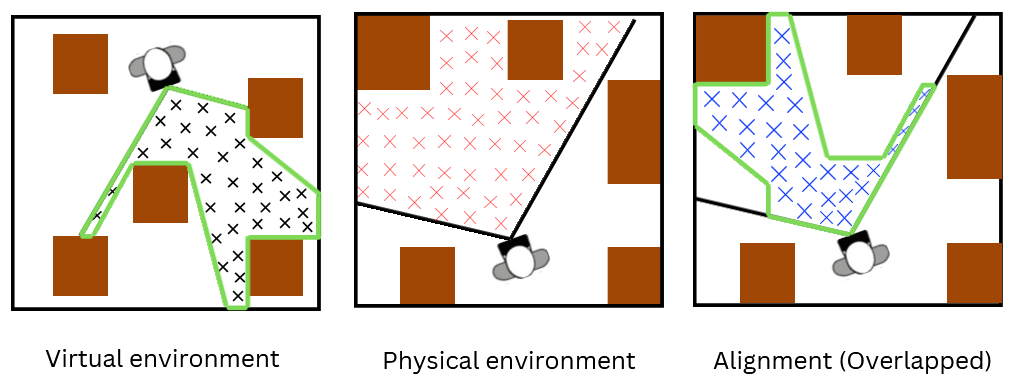}
  \caption{Alignment Target Orientation Calculation Method Example: In this method the alignment is defined by the overlap of open space in view, as such the target orientation is determined by finding what orientation will increase the overlap from the current state.}
  \label{fig:align}
  \vspace{-5mm}
\end{figure*}

To effectively redirect users, the algorithm determines the ideal orientation, preventing collisions and enabling continued uninterrupted walking. Early RDW techniques Steer-to-Center (S2C) and Steer-to-Orbit (S2O), use explicit steering targets, directing users toward the center or a circular orbit around it, respectively~\cite{Razzaque05}.
Bachmann et al. \cite{Bachmann19} was one of the first to use APFs for RDW, introducing an avoidance-based technique using repulsion forces from environmental obstacles to determine the target orientation.
This concept inspired our definition of avoidance-based orientation calculation. However, some non-APF avoidance-based techniques, like Model Predictive Control Redirected Walking (MPCRed), predate APF methods ~\cite{Nescher14}.
Finally, Thomas et al. \cite{Thomas20b} defined environmental alignment between PE and VE, introducing alignment-based redirection, which orients users to maximize an alignment metric.
This alignment metric varies by technique, often including overlapping visible free space or overlapping interaction targets. 
Table~\ref{tab:target} categorizes the corpus by target orientation calculation type used in proposed RDW techniques.
Figures~\ref{fig:steer-and-apf} \&~\ref{fig:align} illustrate principles behind target orientation calculation methods. 
Steering methods direct user locomotion towards one or multiple targets in the PE. Avoidance methods use obstacle-based avoidance forces in the PE. Alignment methods overlap the PE and VE, redirecting users to states that enhance alignment as defined by an alignment metric.

\begin{table*}
  \caption{Classification of Papers by Target Orientation Calculation Type}
  \label{tab:target}
  \centering
  \resizebox{0.9\textwidth}{!}{
  \begin{tabular}{l|l|l}
    \toprule
    Calculation Type & Sources & Percentage of Work\\
    \midrule
    Steering & & 54.55\%\\
    Steer-to-Point(s) & ~\cite{Azmandian22a},~\cite{Azmandian17},~\cite{Bachmann13},~\cite{Cools19},~\cite{Chen17},~\cite{Chen24},~\cite{Dong20},~\cite{Hodgson13},~\cite{Hodgson08},~\cite{Hoshikawa24},~\cite{Joshi20},~\cite{Kwon22},~\cite{Lee19},~\cite{Lee20}, \\
    &~\cite{Mayor22},~\cite{Min20},~\cite{Razzaque05},~\cite{Rewkowski19},~\cite{Schmelter21},~\cite{Sra18},~\cite{Stormer23},~\cite{Thomas19},~\cite{Xu24a},~\cite{Xu24b}\\
    Steer-to-Path & ~\cite{Bachmann13},~\cite{Clarence24},~\cite{Dong19b},~\cite{Fan23b},~\cite{Hodgson13},~\cite{Hodgson14},~\cite{Huang23},~\cite{Kanani17},~\cite{Li20},~\cite{Li23},~\cite{Matsumoto19},~\cite{Neth12},\\
    &~\cite{Qi20},~\cite{Qi23},~\cite{Razzaque05},~\cite{Razzaque01},~\cite{Ropelato22},~\cite{Strauss20},~\cite{Thomas22},~\cite{Xu22b},~\cite{Yu18}\\
    \midrule
    Avoidance & & 29.87\%\\
    APF & ~\cite{Bachmann19},~\cite{Chen24},~\cite{Dong20},~\cite{Dong21},~\cite{Dong23},~\cite{Messinger19},~\cite{Thomas19} \\
    Non-APF & ~\cite{Azmandian22a},~\cite{Chang21},~\cite{Chen21},~\cite{Congdon23},~\cite{Ko20},~\cite{Lee23a},\cite{Lee24c},~\cite{Lee24d},~\cite{Lemic22},~\cite{Liao22},~\cite{Nescher14},~\cite{Sun18},~\cite{Xu22a},\\
    &~\cite{Yang22},~\cite{Zmuda13} & \\ 
    \midrule
    Alignment &~\cite{Chen21},~\cite{KimWoo23},~\cite{Thomas20a},~\cite{Thomas20b},~\cite{Tomar19},~\cite{Wang20},~\cite{Wang24a},~\cite{Wang24b},~\cite{Williams21a},~\cite{Williams21b},~\cite{Wu23},~\cite{Zhang13a} & 15.58\%\\
    \bottomrule
  \end{tabular}
  }
\end{table*}

\subsection{Steering}
\label{sec:steer}

Steering methods direct users toward a target in the PE, using one or more specified target points or paths in the environment. 
S2C guides users toward the center of the environment, helping them avoid obstacles. Continuously moving toward the center in simple environments with only boundary obstacles and infinite possible gains ensures users never reach a boundary~\cite{Razzaque05}. S2C suffers from oscillating gains once the user moves through the center, causing the target orientation to shift behind the user \cite{Razzaque05}. S2O guides users to walk in a circular orbit around the environment center~\cite{Razzaque05}. 
In an ideal scenario with an infinitely straight virtual path and unbound curvature gain, S2O avoids oscillation issues of S2C, allowing continuous walking along the orbit indefinitely~\cite{Razzaque05}. 
Research showed in VEs where users primarily follow straight paths and do not freely change orientation, S2O outperforms S2C~\cite{Hodgson14, Hodgson13, Razzaque05}. 
The zigzag technique, though basic and restricts to following scripted zigzag virtual paths, is the earliest example of path-based steering methods\cite{Razzaque01}.
Hodgson et al. \cite{Hodgson08} presented the first generalized RDW implementation using S2C with live users, proving practical effectiveness beyond simulations.
Steering methods are simple to implement and versatile, but less effective at redirection than avoidance or alignment methods, and harder to apply in multi-user or complex environments.

\subsubsection{Steer to Point}
Methods steering users toward one or more points adjust locomotion to direct accordingly. For one point, passing through it causes a drastic orientation shift as the target moves behind the user. This issue is less pronounced with multiple points. 
Bachmann et al. \cite{Bachmann13} and Azmandian et al. \cite{Azmandian17} enhanced S2C for multi-user setups using collision detection algorithms and triggering resets preventing collisions. 
Joshi et al. \cite{Joshi20} improved S2C with redirection using peripheral vision, saccades, and blinks, leveraging user fixation on tasks. It showed good performance compared to no redirection, but was not tested against other RDW methods, including basic S2C.

Xu et al. \cite{Xu24a} introduced a jump safe RDW method that discretizes the PE into poses to identify poses deemed safe for jumping (safe zones). The method redirects users toward a target pose within the safe zone when approaching VE locations requiring jumps, using the distance the user must travel to the jump site. 
Lee et al. \cite{Lee19} proposed Steer-to-optimal-target (S2OT), employing Q-learning to learn a policy to dynamically determine the optimal steering target. The target updates during redirection, outperforming S2C and MPCRed. 
Lee et al. \cite{Lee20} extended S2OT with Multiuser-Steer-to-Optimal-Target (MS2OT), utilizing Q-learning with a Dueling Double Deep Network structure to select optimal steering targets, predict collisions, and perform pre-reset actions to avoid them. 

Schmelter et al. \cite{Schmelter21} proposed Steer-to-Action (S2A), redirecting users in an active zone only when the next one is unreachable, adjusting their orientation to ensure the next zone becomes reachable. 
Sra et al. \cite{Sra18} presented a RDW method that redirects users only when they exit a smaller 'safe zone' within the tracking space, steering them back into the safe zone. 
Hoshikawa et al. \cite{Hoshikawa24} introduced RedirectedDoors+ (RDD+), involving a robot with a doorknob for haptic door representation. RDD+ combines S2C and S2A, redirecting users toward the environment-center during door interactions. It requires constrained environments designed for this technique. Among steering methods, steer-to-point techniques are simpler than steer-to-path techniques and show greater effectiveness in VEs where user locomotion is less constrained, allowing free direction changes.

\subsubsection{Steer to Path}
These methods steer users to follow a target path closely, rather than walking directly to a specific point.
Since users are moving and not stationary, these methods assume directing them along collision-free paths over directing them to a static point would be more effective.
Kanani et al. \cite{Kanani17} tested S2O-like RDW techniques using hypotrochoid orbits, intending to combine S2C and S2O benefits. They found the four petal rose shape was most effective and preferred by users over S2C and S2O. 
Matsumoto et al. \cite{Matsumoto19} proposed an S2O-like method requiring specially designed VEs with hallways and a special PE, with the use of curved handrails. This haptic method creates the illusion of walking straight while following a curved path, enabling infinite corridor navigation but limited to hallways.
Ropelato et al. \cite{Ropelato22} proposed a steering method designed for maze-like VEs, using large gains to steer users in a straight line across the environment and back repeatedly. High rotations are applied at corners and large translations in corridors. 
Fan et al. \cite{Fan23b} proposed a steering method generating a heatmap of historical walking data to guide redirection. The environment is discretized into poses, steering users to the optimal reachable pose based on the heatmap. This method outperformed predictive algorithms in spaces with internal obstacles and is highly generalizable~\cite{Fan23b}. 
Thomas et al. \cite{Thomas22} suggested using inverse kinematics to determine redirections. The method determines a path between start and finish points entirely within the PE, steering users along a scripted path via pre-calculated gains. 
Xu et al. \cite{Xu22b} proposed a Point-of-Interest (POI) aware steering method that considers reset timing and POI proximity. It triggers early resets if needed to reach the currently selected POI, steering users along the longest possible walkable path to the virtual target. This approach reduces resets and increases reset distance from POIs~\cite{Xu22b}. 

Multiple works presented steering methods that use a Voronoi graph of the VE and a VE-to-PE mapping to generate desired paths to guide users. These methods outperform S2C \cite{Qi20, Li20, Li23} and MPCRed \cite{Li20, Li23}. 
These methods require VEs capable of generating Voronoi maps, limiting their use, but making them ideal for scenarios like guided tours, where such constraints are feasible.
Qi et al. \cite{Qi23} proposed a mapping-based steering method that segments the VE map and calculates the PE path using B\'ezier curves for smooth and reset-free navigation. Nevertheless, curvature gains can exceed DTs and become noticeable. 
Dong et al. \cite{Dong19b} combined environment-based and perception-based RDW by applying an intermediate smooth mapping technique to a larger space than the PE, then using steering-based redirection between the PE and this intermediate mapping. 
Huang et al. \cite{Huang23} proposes Connectivity-Aware Redirected Path Finding (CAW-RPF), a RDW method for wireless VR. It steers users to target vertices using predicted trajectories, obstacle positions, and signal-to-interference-plus-noise ratio (SINR) map. 
Strauss et al. \cite{Strauss20} proposed an RL-based steering method, showing marginal improvements over S2C in some cases, but no significant overall difference. Compared to steer-to-point methods, steer-to-path approaches are more complex to implement and perform better when VE user locomotion is constrained, allowing better mapping to a path within the PE.

\subsection{Avoidance}
\label{sec:avoid}

Avoidance methods use a repulsive force from obstacles in the environment to determine the target orientation for redirection. 
The concept of APF-based RDW methods inspired the definition of avoidance target orientation calculation. 
APFs are vector fields created by summing all force vectors in the field. For RDW, this is achieved by having obstacles generate repulsion forces~\cite{Bachmann19}. The resultant vector in the APF at the user's location determines the target orientation, as it points in the direction of least resistance~\cite{Bachmann19}. However, APFs are just one example of methods using avoidance to determine target orientation. Some planning or learning methods, like MPCRed, typically assign action costs based on likelihood of reset occurrence. These methods minimize this cost, functioning as avoidance-based redirection by assigning orientations to avoid resets~\cite{Nescher14}. Some methods even combine repulsive forces with steering (attractive) forces~\cite{Bachmann19, Dong20}. 
Avoidance methods strike a balance between implementation complexity and performance. While not as effective as alignment methods for specific use cases, they are slightly simpler to implement. 
Avoidance methods generally perform similar to or better than steering methods. In complex PEs or multi-user setups, avoidance methods generally outperform steering methods. 
Moreover, avoidance methods simplify multi-user support by treating other users as sources of avoidance forces, eliminating needing to predict collisions.

\subsubsection{APF}

APF-RDW proposed by Bachmann et al. \cite{Bachmann19} uses repulsive forces from environment boundaries and other users. It also introduces an optional attractive force component, though not used in the basic implementation.
APF-RDW showed 65.6\% fewer resets than S2C in test environments and successfully handled three simultaneous users in a live demonstration without user collisions~\cite{Bachmann19}. 
Messinger et al. \cite{Messinger19} extended APF-RDW by enhancing repulsive forces calculation with proximity scaling. This approach showed improved average performance over standard APF-RDW, particularly in irregularly shaped PEs.
Thomas et al. \cite{Thomas19} proposed P2R, an algorithm using APF-based redirection for complex environments with interior obstacles or non-convex boundaries. In complex PEs, it outperforms S2C, but in simple square PEs, it performs similarly to S2C, as forces effectively guide users to the center, like S2C. 
Dong et al. \cite{Dong20} extended APF-RDW by incorporating an attractive force, combining avoidance with a steering target. This showed 20\% improvement over APF-RDW in most cases.
Dong et al. \cite{Dong21} proposed Dynamic Density-Based Redirected Walking (DDB-RDW), a multi-user RDW method dividing the PE into concentric density zones. Higher density at the center acts as a repulsive or attractive force to distribute users among zones, combined with APF to avoid collisions with boundaries, reducing resets and distance to center compared to APF-RDW.
Chen et al. \cite{Chen24} proposed APF-S2T, combining APFs, steering targets, and visibility polygons. Visibility polygons represent the free area within users' view. By intersecting PE and VE visibility polygons and forces from APFs, the algorithm identifies a steering target within this reachable area and directs the user toward it. 
This method outperformed APF-RDW in complex PEs containing numerous internal obstacles~\cite{Chen24}. Amongst avoidance methods, APFs provide simplest implementation while offering strong performance across most use-cases.

\subsubsection{Non-APF}
Zmuda et al. \cite{Zmuda13} proposed Fully Optimized Redirected Walking for Constrained Environments (FORCE), which predicts users' future movements and performs redirections by evaluating possible end states to redirect users toward more favorable positions, reducing likelihood of future resets. 
FORCE shows 10\% improvement over S2C in constrained environments; however, FORCE is only applicable in constrained environments where future movement is limited~\cite{Zmuda13}. 
Nescher et al. \cite{Nescher14} proposed MPCRed, a planning algorithm that calculates target orientation based on the action most likely to avoid collisions from possible future paths.
Ko et al. \cite{Ko20} suggested using deep RL to learn a policy to apply redirections. The reward function penalizes resets and proximity to boundaries. This showed improved performance over S2C. 
Chang et al. \cite{Chang21} proposes using RL to learn a policy using similar rewards penalizing resets and proximity to boundaries. This showed improved performance over non-RL controllers, and exhibited behavior oscillating between S2C and S2O. 
Xu et al. \cite{Xu22a} contributed a method that discretizes the PE into standard poses. Users are redirected to the safest reachable standard pose. Safety representing the chance of avoiding an obstacle.

Sun et al. \cite{Sun18} proposed Dynamic Saccadic Redirection, which leverages decreased sensitivity to redirection during saccades. The algorithm detects saccades and uses real-time path planning to redirect users during these saccades. Target orientation is calculated using both static boundary avoidance and dynamic obstacle avoidance factors. The method outperformed S2C in simulations~\cite{Sun18}. 
Liao et al. \cite{Liao22} introduced a path planning algorithm that minimizes accumulated cybersickness from gains and SINR penalties. It calculates a graph of the environment considering cybersickness costs and SINR quality, avoiding obstacles that block line-of-sight (LOS) signals. The method showed reduced cybersickness and improved SINR ratings compared to other approaches. 
Congdon et al. \cite{Congdon23} proposed Monte-Carlo Redirected Walking (MCRDW), applying the Monte-Carlo method to RDW. Recognizing that ideal redirection strategies may shift as users explore, the method scores strategies based on boundary avoidance and gain subtlety, applying the highest scoring redirection while pushing scores to a queue. MCRDW reduced boundary collisions by 50\% compared to S2C~\cite{Congdon23}. 

Yang et al. \cite{Yang22} proposed a portable RDW method utilizing inertial and laser ranging sensors. Unlike traditional RDW methods requiring a known fixed PE, this approach does not rely on PE foreknowledge and instead redirects based on sensor information to avoid collisions. They demonstrate that the method safely redirects users without prior knowledge of the PE.
Lee et al. \cite{Lee23a} suggested user-centered RDW (UC-RDW) using collision sensors that emit 'virtual feelers' to detect obstacles within the environment, and generates repulsive forces based on obstacle proximity to calculate target orientation. UC-RDW outperformed S2C and S2O, but underperformed compared to APF-RDW~\cite{Lee23a}. 
These two methods are noteworthy for using sensors to determine redirection, enabling greater portability by eliminating the need for predefined PEs.

\subsection{Alignment}
\label{sec:alignment}
The concept of alignment between PEs and VEs was proposed by Thomas et al. \cite{Thomas20b}. These alignment methods define a metric for environmental alignment, some examples of this metric are: the overlap of interaction targets in the PE and VE for passive haptics experiences, or the overlap of visible walkable free-space in the PE and VE from current view. The methods then determine the target orientation towards whichever orientation will optimize this alignment metric. Alignment-based techniques improve passive haptic experiences by aligning both obstacles in the physical and VEs and haptic targets in the PE with interaction points in the VE, this is done such that when a user arrives at an interaction point in the VE they will also arrive at a haptic target within the PE~\cite{Thomas20b}. Thomas et al. \cite{Thomas20b} propose an enhancement to P2R using APFs with both repulsive and attractive forces to maintain this concept of alignment, it differs from just using APFs with steering targets as it contains a mathematical foundation of the alignment conditions. They showed that this concept of alignment can enable RDW techniques to support physical interactivity~\cite{Thomas20b}. 
Thomas et al. \cite{Thomas20a} propose a mathematical foundation to generalize environmental alignment and verify with simulation this mathematical foundation. They show switching between avoidance and alignment as more effective than always using both. From this they suggest that a controller may be more effective by designing when to switch to alignment-only redirection \cite{Thomas20a}.

Alignment shows the best performance of all target orientation calculation types when environmental alignment is imperative to the function of the application (such as for passive haptic interactions), alignment methods are the most complicated of the heading calculation types to implement and as shown by \cite{Thomas20a}, there are times that avoidance can outperform alignment even within the same environment where at other times alignment can outperform avoidance. Currently, alignment shows the best overall performance, though it is also the most complicated target orientation calculation method to implement.

Before Thomas et al. \cite{Thomas20b} proposed the concept of alignment for RDW, Tomar et al. \cite{Tomar19} proposed a RDW controller for remote telepresence that we have classified as an alignment based algorithm: it takes both the local PE and remote PE and maps them to a canonical space such that redirection occurs with respect to this canonical space to afford remote telepresence by deforming both spaces. Kim et al. \cite{KimWoo23} propose a similar concept of spatial alignment between local and remote environments using relative translation gains (RTGs): discretizing the space into a RTG grid allowing pairs of RTGs to be applied to the grid. They found increased performance over other RTG space matching techniques and defined new metrics for environment similarity and complexity for evaluating environment mapping techniques~\cite{KimWoo23}.

Wang et al. \cite{Wang20} propose a RDW method for a constrained VE that dynamically aligns users with haptic targets using b\'ezier curve path transformation, they showed good user acceptance and improved presence as compared to other tested RDW methods. Williams et al. \cite{Williams21a} propose Alignment-based redirection controller (ARC), which expands upon the definition of alignment to redirect users to maintain environment similarity as defined by their similarity metric, taking into account proximity to obstacles in the VE and PE. ARC unlike other alignment controllers is primarily used for collision avoidance rather than passive haptic support \cite{Williams21a}. They found ARC outperformed S2C and P2R for number of resets and distance walked, and that it applied less intense curvature gains~\cite{Williams21a}. Williams et al. \cite{Williams21b} propose using the concept of visibility polygons to determine orientation by superimposing free spaces from both the PE and VE to redirect users to best align the visible free space from the PE and VE. They found that the algorithm outperformed ARC in static scenes and that in scenes with dynamic virtual obstacles it significantly outperformed ARC~\cite{Williams21b}. Wu et al. \cite{Wu23} proposes a RDW method combining the visibility polygons free space alignment idea with APFs. They found the alignment only and alignment plus APF controllers to outperform Williams et al. \cite{Williams21b}'s visibility polygon controller \cite{Wu23}. Wang et al. \cite{Wang24a} proposes Transferable Virtual-physical Environmental Alignment (TRAVEL) a RL based alignment controller that performs dynamic matching of haptic targets, redirecting user to avoid obstacles and maintain alignment with interaction targets, showing improved performance over other alignment algorithms, though they acknowledge that the dynamic target matching may break down if physical targets are not sparse enough. Wang et al. \cite{Wang24b} proposes combining the concepts of RDW with omni-directional treadmills, they leverage the ability of RDW alignment controllers to support passive haptics while also aligning users with the direction of the omni-directional treadmill with higher speed range. They show that the method successfully combines RDW and omni-directional treadmills to leverage the benefits of RDW for improved UX by aligning users with the optimal treadmill direction and enabling haptic interaction with haptic targets located around the treadmill~\cite{Wang24b}. This study shows the potential for RDW to be used beyond the original design and the potential that remains to be explored.

\section{Gain Application Calculation}
\label{sec:alg_type}

Table~\ref{tab:redirtype} shows the papers organized by the gain application type used in the RDW method(s) within the paper. As most of the work was discussed in Section~\ref{sec:heading_calc} with relation to \textit{Target Orientation Calculation} type we give a brief overview of what these gain application types are and how the works fit into these categorizations without going into as much detail on the methods except where they add something specifically important for \textit{Gain Application}.

\begin{table*}
  \caption{Classification of Papers by Gain Application Calculation Type}
  \label{tab:redirtype}
  \centering
  \resizebox{0.9\textwidth}{!}{
  \begin{tabular}{l|l|l}
    \toprule
    Calculation Type & Sources& Percentage of Work\\
    \midrule
    Reactive &~\cite{Azmandian17},~\cite{Azmandian22a},~\cite{Bachmann19},~\cite{Bachmann13},~\cite{Chang21},~\cite{Chen17},~\cite{Chen24},~\cite{Chen21},~\cite{Clarence24},~\cite{Cools19},~\cite{Dong20},~\cite{Dong23},~\cite{Dong21}, & 63.16\%\\
    &~\cite{Hodgson13},~\cite{Hodgson14},~\cite{Hodgson08},~\cite{Hoshikawa24},~\cite{Joshi20},~\cite{Kanani17},~\cite{Ko20},~\cite{Lee23a},~\cite{Lee24c},~\cite{Lee24d},~\cite{Messinger19},~\cite{Min20}, \\
    &~\cite{Neth12},~\cite{Razzaque05},~\cite{Rewkowski19},~\cite{Schmelter21},~\cite{Sra18},~\cite{Stormer23},~\cite{Strauss20},~\cite{Sun18},~\cite{Thomas20a},~\cite{Thomas19},\\
    &~\cite{Thomas20b},~\cite{Wang20},~\cite{Wang24a},~\cite{Wang24b},~\cite{Williams21a},~\cite{Williams21b},~\cite{Wu23},~\cite{Xu24a},~\cite{Xu22b},~\cite{Xu22a},\\
    &~\cite{Yang22},~\cite{Yu18},~\cite{Zhang13a}\\
    \midrule
    Predictive &~\cite{Huang23} & 18.42\%\\
    Graph Prediction & ~\cite{Azmandian22a},~\cite{Congdon23},~\cite{Fan23b},~\cite{Nescher14},~\cite{Zank17},~\cite{Zmuda13} &\\
    Heuristic Prediction &~\cite{Li24},~\cite{Mayor22},~\cite{Xu24b} &\\
    Reinforcement Learning &~\cite{Jeon24},~\cite{Lee19},~\cite{Lee20},~\cite{Lemic22} &\\
    \midrule
    Scripted & & 18.42\%\\
    Pre-Calculated Path&~\cite{Azmandian22a},~\cite{Liao22},~\cite{Qi20},~\cite{Qi23},~\cite{Razzaque01} & \\
    Environmental Mapping&~\cite{Dong19b},~\cite{KimWoo23},~\cite{Kwon22},~\cite{Li20},~\cite{Li23},~\cite{Matsumoto19},~\cite{Ropelato22},~\cite{Thomas22},~\cite{Tomar19}\\
    \bottomrule
  \end{tabular}
  }
  \vspace{-4mm}
\end{table*}

\subsection{Reactive}
\label{sec:react}

Reactive algorithms do not consider any future movements of the user, instead they apply gains towards the currently calculated desired heading from the user's current position, regardless of the user’s future movements. Reactive algorithms make the bulk of RDW algorithms, and the first algorithms proposed were reactive algorithms. S2C, S2O, and APF-RDW are all prominent reactive algorithms \cite{Bachmann19, Razzaque05}. For each reactive algorithm, first the algorithm determines the desired heading using their respective desired heading calculation technique and then applies gains to the current view frame based on the current state to achieve redirection towards the desired heading. Table~\ref{tab:redirtype} shows all the papers focused on reactive algorithms in the data set, these papers were discussed early in reference to their desired heading calculation. 

One of the strengths of reactive algorithms is that they can be designed to be a general solution RDW algorithm that can handle any VE, whereas predictive or scripted algorithms may instead require constraints on the VE. Reactive algorithms make no assumption or constraint on direction of user locomotion, allowing it to freely respond to any changes in direction users may make, though since it responds after the user moves and makes no assumption about future movement it may fail to optimally redirect users. Consider the case where a user intends to rotate 180$^{\circ}$, since reactive algorithms respond per frame update and are unaware of the user's intention to rotate 180$^{\circ}$ the gains applied to move the rotation towards the center for S2C could result in shrinking the rotation in the PE that causes the user to be directed further from the center than had the algorithm allowed some of the rotation to be un-modified or had the rotation been expanded, however the reactive algorithm had no way of knowing the final orientation of the user for the rotation and could only respond frame by frame after a section of user rotation. Reactive RDW is simple, generalizable, easy to implement, and still shows good performance despite possible downsides compared to future-aware redirection.

\subsection{Predictive}
\label{sec:pred}
Predictive algorithms make a prediction of future locomotion and calculate and apply gains based on this future movement. Leveraging this predicted future movement allows predictive techniques to consider the future state of the user after redirection based upon this expected movement to better direct the future state of the user than just manipulating based on their current heading. There are a few different categories of predictive methods. There is the classical method of using a constrained environment with a simple graph based prediction that simplifies the prediction to either walking straight down a hallway or an equal probability split of taking a fork at an intersection. There is the heuristic prediction method that generates a discretized simplified list of possible paths, trimming similar paths into one representative path, and uses heuristic information to determine likelihood of paths being taken such as where a goal in the VE is. There are also RL based locomotion predictor methods, these methods are fewer due to their complexity though they do not require environmental constraints. CAW-RPF relies on predictions of future path, but does not propose a prediction method, the method requires a singular predicted path as input and outputs a planned path considering connectivity constrains \cite{Huang23}. CAW-RPF could use any of the prediction methods used for predictive RDW provided it is provided with a single predicted path.

\subsubsection{Graph Prediction}
Graph-based prediction methods rely on a constrained VE with limited locomotion choices from which a simplified skeleton graph of the environment can be constructed. This is often done using a Voronoi graph, from here the methods use this graph of the environment to make a prediction about user locomotion using the graph and additional state information sometimes including heading, virtual targets, or gaze direction. These prediction methods output a prediction of all possible paths the user could take under the simplified environment graph with corresponding probabilities of being taken: at times some methods may produce only one path if the user is not at an intersection \cite{Congdon23, Fan23b, Nescher14, Zank17, Zmuda13}. Once the methods have a prediction for the user's locomotion, the gains are applied to this predicted path so the resultant path leads towards the target orientation in the PE as calculated by the respective calculation method. When producing multiple path options, the methods will weigh all actions against all predicted paths, choosing the action that is the best given all possible future paths with weighting of path importance to cost of an action applied by path probability \cite{Congdon23, Fan23b, Nescher14, Zmuda13}.

Zank et al. \cite{Zank17} proposes a graph extraction method and path prediction method using this graph that can be used by RDW, this method does so by getting the mesh of the VE, then calculating a Voronoi graph from this, and using this Voronoi graph to produce a skeleton graph of the environment, then online predictions are made using visibility polygons to determine what vertices of the walk-able mesh calculated before are visible in a 360$^{\circ}$ field of view from the user's current location, this information is then used to generate possible future paths, then eye-tracking is used to determine the likelihood of possible paths, this prediction can then be fed to a RDW method such as MPCRed. This method assumes a static VE, but they propose that many dynamic objects in virtual scenes could be ignored as doors can be opened and interaction targets moved by users will not be important to the mesh \cite{Zank17}. 

\subsubsection{Heuristic Predictors}
Heuristic-based prediction methods are used to make a decent assumption of future locomotion that can be exploited. Some methods use mathematical formulas to make predictions of short-term locomotion based on user trends and current heading \cite{Mayor22}, and others assume the user will walk towards a virtual target \cite{Xu24b}. Once the prediction about user locomotion is obtained, the method applies gains to the predicted path to direct PE locomotion towards the target orientation.

\subsubsection{Reinforcement Learning Predictors}
RL has also seen use as a prediction method by constructing locomotion predictors to predict short-term or long-term locomotion of users \cite{Jeon24, Lee19, Lee20, Lemic22}. Once these predicted paths are determined the gains can be applied to the predicted path such that the path in the PE will bring users to the desired heading. RL based prediction methods do not always require a constrained VE and can operate in an open VE with more locomotion opportunities since they do not rely on environmental simplifications to make these predictions \cite{Jeon24, Lee19, Lee20, Lemic22}. 
Lemic et al. \cite{Lemic22} proposes using RL with Recurrent Neural Networks (RNNs) to predict locomotion for RDW and feeds these predictions APF-RDW to perform redirection, they show a proof of concept simulation that shows these RNNs with near future prediction can function for RDW. Jeon et al. \cite{Jeon24} proposes F-RDW, a RL-based prediction method for RDW that feeds these predictions into other RDW methods such as MPCRed and S2C to then calculate the gains to apply to the predicted path, they saw that this method was able to improve the performance of the RDW methods it was applied to such as MPCRed and S2C for all tested environments including empty environments and complex environments showing a statistically significant improvement in small environments.

\subsection{Scripted}
\label{sec:script}
Scripted algorithms often work similarly to graph-based prediction methods, however instead of making online predictions about user locomotion then dynamically applying gains, these methods rely on offline pre-calculated gains for a pre-scripted path for the user to follow such as for guided tours or a pre-calculated mapping from the VE to the PE, both these methods require pre-processing with both the VE and PE known. Since the method relies on offline pre-processing it requires a VE of fixed size and shape, often with restraints such as a basic maze, to simplify the processing using skeleton graphs.

\subsubsection{Pre-calculated Path}
One example of scripted RDW methods is using a pre-set target location for the user and using path planning methods to pre-calculate the path with gains in the PE to produce the least amount of resets for the user, these methods often involve using a graph representation of the VE to accomplish the path planning \cite{Ko19, Liao22, Qi20, Qi23, Razzaque01}. Some of these methods use a path planning component that selects between VE paths when there are multiple options available so as to produce both the virtual path and gains such that the resultant physical path has the fewest resets \cite{Ko19, Qi20}, some also consider additional information such as SINR strength or cybersickness costs \cite{Liao22}.

Ko et al. \cite{Ko19} redefine RDW as a graph search problem, defining both the physical and virtual world as graphs and given a starting point and destination point defines an approximation scheme to find a solution to the search for a path that solves the graph of the virtual path and physical path with gains \cite{Ko19}. However, this method relies on users following the specified path and also relies on an input of the destination, while this could be fed with a predicted destination path to be more predictive, it is still a path planning algorithm that plans both virtual and physical paths so relies on users following the planned virtual path, as such it fits more into this category of scripted RDW since it relies on a predicted path with no branches and functions only if users follow this plan.

\subsubsection{Environmental Mapping}
These methods frame RDW as a planning problem, they propose making a mapping from the VE to PE such that at any point the gains that should be applied to result in paths with the fewest resets are known. Many of these methods, similarly to graph based predictive methods, require a constrained environment from which a Voronoi map or skeleton graph can be created, then using this map of the VE, these methods then calculate gains to apply by creating a mapping from the path segments in the VE to the PE \cite{Dong19b, Li20, Li23, Matsumoto19}. Overall, these methods do exploit similar things as the graph predictive methods and the main difference is that the mapping for the virtual paths to the PE is pre-calculated. The main downside of these methods is that they are not applying dynamic gains; therefore, if small deviations in user locomotion add up users can be redirected less effectively, or if users take odd or unexpected paths in the VE they could result in less desirable paths in the PE depending on how the path segments they do take have been mapped to the PE.

Other mapping methods divide the environments into a grid for mapping, typically for re-scaling for remote collaboration, these methods then map the spaces together using some method, such as aligning edges of the environments using the grid representation to determine gains to achieve this alignment \cite{KimWoo23, Tomar19}. These techniques are similar to the graph based mapping but are used in environments of similar shape such as two rectangular rooms for remote collaboration, so have different considerations and limitations. A similar method is using a series of hallways or tunnels that are mapped such that the PE is scaled with translation gains to map to the entire hallway segment, and once the user reaches a fork then rotation gains are applied when the user changes direction so as to map the PE to the new hallway or tunnel segment of the VE \cite{Ropelato22}.

Kwon et al. \cite{Kwon22} propose a method that relies on spacial tiling and overt reorientations at fixed positions rather than subtle redirections. They show that this method can take a given VE and PE and perform a mapping between the two such that the user is prompted to stop and reset only at fixed positions and is otherwise free to explore the current tile until they move to a reset point, highlighted in their view, to initiate a reorientation and move to the next tile. They found that users reported higher sense of presence using this technique than APF-RDW for an observation-based task.


\section{General Enhancements}
\label{sec:enhancements}

Some work includes general enhancements to RDW techniques that can be applied to any existing technique to improve performance. These enhancements currently fall into the categories of multi-user support, complex PE support, gain masking techniques, expanded motion types, and immersion improvements. These enhancements are optional and adaptable to any RDW algorithm. This section outlines enhancement categories found in our archive but is not an exhaustive list. 

\subsection{Multi-User}
Initial RDW methods supported only one user, however research has been done to find ways to extend RDW methods to support multiple simultaneous users. 
Early RDW methods supported only a single user, but research explored extending them to accommodate multiple simultaneous users. Bachmann et al. \cite{Bachmann13} enhanced S2C with collision detection and avoidance methods, which can be applied to other single-user RDW methods to enable multi-user support. 
Azmandian et al. \cite{Azmandian17} showed that in smaller physical spaces, shared usage is more effective than subdivision. They found that heuristic conditions for user-user collision avoidance were not fully reliable but improved when combining proximity, relative velocity, and a reset condition to halt both users when necessary.
Dong et al. \cite{Dong19a} proposed a collision detection and avoidance method for three co-located users, extending previous methods that supported only two users. Their simulations achieved a 90\% collision avoidance rate.
However, testing with real users poses safety risks due to the 90\% collision avoidance rate, though it marks progress toward multi-user support beyond two users.
Similarly MS2OT predicts user collisions and prevents them with a pre-reset action~\cite{Lee20}. This approach, leveraging RL, is more closely tied into the method, making it less applicable to other methods, though similar techniques could be applied to other RL controllers. 
Lemic et al. \cite{Lemic22} proposed two Recurrent Neural Network (RNN) types for training RL-based locomotion prediction in RDW methods. Their approach demonstrated effective locomotion prediction, even when applied to different numbers of users than those used during training.

Another strategy for multi-user support is space division, where each user is assigned a subdivision of the space. The RDW method is applied independently within each subdivision, using virtual barriers to maintain separation. To improve upon static space sharing methods, Jeon et al. \cite{Jeon22} proposed RL-based dynamic space sharing, which adjusts virtual barriers dynamically for optimal space sharing between users, outperforming static equal split strategies. 
Lee et al. \cite{Lee24c} introduced the Multi-Agent Reinforcement Reseter (MARR), a multi-user reset technique for RDW. Using a Markov Decision Process and multi-agent reinforcement learning, MARR models users collaborating to minimize resets by optimizing reset directions. Their reward function incentivizes distance from obstacles and penalizes resets.
They show that MARR significantly reduces resets compared to other techniques, particularly in complex environments~\cite{Lee24c}.
Dong et al. \cite{Dong19b} introduced a user-controlled collision avoidance and interaction method eliminating the need for resets or redirection. By rendering avatar representations between users' real positions, the method alerts users to others' presence, enabling them to avoid collisions or initiate interactions. 
Min et al. \cite{Min20} proposed a recovery algorithm for RDW applications to enable multi-user collaboration by aligning user's relative virtual and physical locations. Unlike many methods that focus solely on user avoidance, this approach addresses scenarios where co-located users in the VE are misaligned in the PE, which can hinder collaboration and feel unnatural. The recovery method corrects this misalignment to support seamless collaboration.

APF RDW methods support multiple users by treating other users as dynamic obstacles exerting avoidance forces on other users~\cite{Bachmann19, Messinger19}. 
APF-RDW is able to support up to eight concurrent users with significantly fewer resets than no redirection, but as the number of users increase the resets increase~\cite{Bachmann19}.
Dong et al. \cite{Dong20} adds enhancements to APF-RDW using a dynamic steering target per user and use repulsion forces from higher-redirection-priority users predicted next location for other users redirection. They saw a 20\% reduction of resets over APF-RDW \cite{Dong20}.
DDB-RDW uses APFs with density based clustering zones for multi-user RDW by distancing users with APF and distributing users among density zones so no zone contains too high a density of users, showing improvements over APF-RDW with higher number of concurrent users with increased benefits in larger PEs~\cite{Dong21}. 

\subsubsection{Wireless Communication Multi-player}
VR users may stream games wirelessly from PCs to headsets or, in multi-user applications, connect to a central base station streaming the application to all users. 
Research explored RDW modifications that handle the unique challenges and leverage benefits of wireless communications. 
Van Onsem et al. \cite{VanOnsem23} propose multiple improvements: latency-aware modifications to handle data transmission delays, a drifting threshold to discard minor positional changes and avoid uncomfortable gains, and a redirection history component to detect and reduce oscillations that degrade user experience and performance.
These enhancements improved performance in both single- and multi-user settings~\cite{VanOnsem23}. Similarly, CAW-RDW enhances RDW with wireless-aware modifications, prioritizing a stable connection between the base station and user HMD~\cite{Huang23}.

\subsubsection{Remote Multi-player}
Multi-user RDW typically focuses on supporting co-located users, but, accommodating remote multi-user applications is important. 
Any RDW algorithm can be applied to remote users without user collision concerns, but these methods overlook other concerns like multi-player reset fairness or spacial consistency between remote environments. 
Tomar et al. \cite{Tomar19} suggested a method that applies redirection to distort remote spaces, supporting spacial consistency. This improves support for interactions between remote users.
Fairness in multi-player applications becomes a concern when RDW requires resets, as one user may gain an advantage while another user resets. This is especially critical in competitive scenarios (e.g. first person shooter game).
Xu et al. \cite{Xu24b} introduced synchronized resets to improve fairness in competitive multi-user applications using RDW. 
They propose that when one user requires a reset, the system should simultaneously reset all users. This maintains fairness by temporarily removing all players from the game while resetting~\cite{Xu24b}. 
This issue is exacerbated with remote users, as differing PE sizes can cause some users to experience more resets and interruptions than others.

\subsection{Expanding RDW Motions}
Traditionally, RDW is restricted to forward walking along a 2D plane, with $y$ as the vertical direction. Gains are applied only to movements on $x, z$ plane.
Recent advancements expanded RDW beyond simple forward movements on a 2D plane. These include Redirected Jumping (RDJ) for redirection during jumps~\cite{Hayashi19,Jung19, Xu24a}, methods for modifying non-forward steps on the 2D plane~\cite{Dong23}, manipulations in roll and pitch directions~\cite{Yamamoto18}, and slope redirection~\cite{Hu19}. 
Pitch, roll, and slope gains have seen limited use in RDW and do not demonstrate the same potential as jumping for greater user redirection with reduced detection.

\subsubsection{Jumping}
\textbf{RDJ} extends RDW beyond forward walking on a 2D plane, following DT values for jumping gains from prior work \cite{Hayashi19,Jung19}. 
Xu et al. \cite{Xu24a}'s SafeRDW algorithm incorporates jumping motions and, while it did not minimize resets, it achieved the highest jump safety levels. 
SafeRDW highlights the potential of jump-aware RDW for applications where jumping is integral, such as games involving obstacle jumping in the VE. It demonstrates potential to harness greater redirection achievable through jumping.

\subsubsection{Non-Forward Motion}
Traditional RDW assumes motion only occurs in user gaze direction, but this does not fully reflect human locomotion. 
While sidestepping, back-stepping, or looking around while walking human movement direction differs from their orientation. 
This can lead to unsafe motion with RDW, as redirections to avoid obstacles assume forward movement. 
FREE-RDW enables non-forward movement by predicting and redirecting sidesteps and back-steps using gains within thresholds calculated by step type (forward, backward, left, right). It resulted in fewer resets for non-forward steps than algorithms not accounting for such movements~\cite{Dong23}. 
This highlights the potential to further expand RDW for applications where non-forwards steps are common, such as fighting games, first-person shooters, or other action-based systems.

\subsection{Gain Masking Techniques}
\subsubsection{Distractors}
Distractor events, events that direct the user's attention, can enhance the opportunities to apply gains to users by encouraging head rotations and mask gains by directing the user's attention to the event~\cite{Cools19, Sra18}.
Chen et al. \cite{Chen17} designed a game using a distractor event when the user leaves an inner "safe circle" zone of the PE, using a dragon the users need to shoot and avoid attacks from to induce more rotations and mask gains applied during the interaction with the dragon. They found continuous S2C combined with the distractor effective at redirecting users. 
S2A uses interaction actions as distractors, applying redirection only while users are in action zones, allowing redirection while users engage in actions~\cite{Schmelter21}.
S2A requires specially designed VR application as it relies on defined action zones as the only source of redirection.  Stormer et al. \cite{Stormer23} used S2A in a multi-user application where users had their own adjacent tracking spaces, they found that collaboration can act as a distractor and induce redirection.

\subsubsection{Saccades}
As mentioned in Section~\ref{sec:gain_factors}, users can be redirected more during saccades and blinks~\cite{Alsaeedi21, Bolte15, Joshi20, Keyvanara18, Keyvanara19, Langbehn18b, Nguyen18c, Sun18}. Some RDW methods leverage this phenomenon applying stronger redirection during these actions, showing improved effectiveness~\cite{Joshi20}. Some methods also induce more saccades to increase redirection opportunities, but induced saccades can feel tiring and unnatural~\cite{Sun18}. Adding this concept to other techniques could improve effectiveness of RDW.

\subsubsection{Non-visual Stimuli}
As mentioned in Section~\ref{sec:gain_factors} non-visual stimuli and haptic devices can mask gain detection during RDW. 
Haptic walls or rails during motion reduce the detection of gains, but this is underutilized in RDW methods and instead remains mostly theoretical, primarily used in research on gain DTs~\cite{Matsumoto16, Matsumoto19, Nakamura19}. 
Lee et al. \cite{Lee24b} proposes a haptic device attached to the hand to extend the concept of wall haptics, they show that this can reduce the detection of curvature gains while walking near a wall by simulating the feeling of wall contact.
Hanger Reflex (HR), the human reflex to involuntarily walk toward a side of the body being compressed, using devices attached to users waists that apply compression on either side influences walking during VR and complements RDW to further distort walking paths of users~\cite{Xie19}. 

Research has shown the use of auditory and olfactory stimuli for redirection, such as sounds to direct user attention or the redirection of the sound of a river to follow, improves RDW applicability ~\cite{Lee24a, Rewkowski19, Weller22}. Research also shows that, in applications with noisy visual stimuli, auditory stimuli can redirect users stronger than visual redirection~\cite{Lee24a, Rewkowski19, Weller22}. The potential for these for RDW is still unexplored as research explored the effect on gain DTs, but few RDW techniques have incorporated auditory or olfactory stimuli. 
Rewkowski et al. \cite{Rewkowski19} tested S2C with gains from auditory stimuli for visually impaired users, and found it less effective due to limited head rotations from reduced looking around, but still promising for the visually impaired. 
Weller et al. \cite{Weller22} used an "auditory path" in an environment with noisy visual information to notify the user when on path to the target, the path was represented with footstep sounds corresponding to being on the path, they found using these auditory cues successfully redirected users and masked the applied gains.

Electrical muscle stimulation shows potential to redirect user steps when combined with traditional visual gain based manipulations~\cite{Auda19}, though has only been tested with users walking straight with either no gains or continuous left rotation. Galvanic Vestibular Stimulation (GVS), non-electrical vestibular stimulation with Bone-Conduction Vibration (BCV), and Caloric Vestibular Stimulation (CVS) influences users' motion and detection of redirection gains~\cite{Hwang23a, Hwang23b, Matsumoto21}. These methods were not implemented into a RDW system to compare performance with state-of-the-art techniques.

\vspace{-3mm}
\subsection{Irregular and Complex Environment}
\label{sec:irreg}
Traditional RDW techniques assume a static square or rectangular PE, the user is able to safely walk throughout this space without any risk of collision~\cite{Razzaque05}. In VR applications, especially when developers cannot control the PE, this may not be the case. 
Some environments have internal obstacles, other environments are not rectangular, such as being L-shaped, traditional RDW techniques such as S2C struggle in these environments.
APF based techniques and alignment techniques support irregular environments by using avoidance forces from obstacles to guide redirection instead of a set steering target~\cite{Chen24, Messinger19, Thomas19, Wang24a, Williams21a, Williams21b, Wu23}.
Xu et al. \cite{Xu22a}'s pose-safety-based avoidance method shows competitive performance in complex environments with internal obstacles.
RL based methods can handle complex environments, these methods are trained in different complex environments and learn a policy to avoid resets or steer along an optimal path for maximum walking distance~\cite{Lee20, Lee24c, Zhang23b}.
Overall, there has been a lot of work on complex environments, but as environments become more complex RDW becomes more difficult. Therefore, it remains important to continue to improve on techniques that can handle complex environments. Dynamic obstacles, although handled by some techniques, are still an under-considered part of complex environments that could see more support, the difficulty is that they require some kind of tracking.
Some methods use sensors on the user, other methods use a base station and the LOS signal strength or SINR strength to detect obstacles, these methods can handle dynamic obstacles~\cite{Lee23a, Liao22, Yang22}.

\subsection{Immersion Improvements}
Immersion improvements enhance RDW by making it integral to the application rather than a separate overlay. 
We define immersion improvements as enhancements that integrate RDW into the system, such as incorporating redirection into the narrative, and improving passive haptic interactions to enhance users' sense of engagement with the VE. 
Freitag et al. \cite{Freitag14} proposed a user-triggered teleportation reset, framing resets as an interaction/locomotion technique rather than a temporary safety interruption. Users select a teleport destination, and a portal appears for users to walk through. The portal is positioned to reorient users toward an optimal heading without applying rotation gains \cite{Freitag14}. 
Since portals served as an interaction technique that users could initiate anytime, not just during required resets, experiments showed some users would teleport near the target as soon as they realized it was unreachable by walking within the PE \cite{Freitag14}. This demonstrates the value of early resets.
Overall, while it allows users to avoid real walking for locomotion, it demonstrates potential as alternative to traditional reset techniques. 
Liu et al. \cite{Liu18} suggested a teleportation reset method designed as a standalone approach separate from RDW. It showcases the potential of engaging reset techniques by letting users specify teleport locations before creating portals they walk through, reorientating them as they turn to walk through portals~\cite{Liu18}. 
These methods could complement RDW, providing a safe and interactive reorientation method for users.  
Yu et al. \cite{Yu18} designed a virtual tour using RDW in a very small PE and a VE of tunnel regions. They used a narrative-driven reset technique, where users reached unlit tunnel sections and had to search for a light switch to proceed. This search induced extra rotations during which they could be redirected to a safe orientation. 
This illustrates how reset techniques can leverage narrative elements to blend seamlessly into the VE, making resets less noticeable to users than traditional stop and turn resets. 
This example relies on a scripted tour to integrate resets into the environment, but there is potential to develop more generalized techniques that build upon this concept.

\subsubsection{Passive Haptics}
Some research currently explored leveraging RDW's control over user orientation to improve passive haptics. It guides users toward haptic targets in the PE as they approach interaction points in the VE. 
Many techniques supporting passive haptics are alignment techniques (See Sec.~\ref{sec:alignment}), however not only those methods support passive haptics. 
Alignment techniques improve passive haptics by steering users toward PE haptic targets as they approach VE interaction points~\cite{Chen21, Thomas20a, Thomas20b, Wang20, Wang24a, Williams21a}.
Clarence et al. \cite{Clarence24} suggested combining RDW with haptic re-targeting techniques to redirect hand movements, aligning haptic targets in the VE with those in the PE during interaction. 
They use RDW similar to aforementioned alignment methods. If RDW alignment is insufficient upon reaching the target, haptic re-targeting adjusts the virtual hand to align with the interaction point while redirecting the physical hand to the physical target, enhancing reachability \cite{Clarence24}. 
RDD+ shows how haptic door interactions can be implemented with and enhance RDW \cite{Hoshikawa24}. 

\section{General Discussion}

Ongoing improvements simplify implementation while enhancing integration benefits of RDW. Multi-user RDW supports co-located users and fair reset methods for remote multiplayer VR games. Enhanced gain masking techniques enable greater redirection in smaller spaces using higher-magnitude gains. Environmental alignment affords improved haptic experiences, sense of presence, granting new avenues for VR applications. Simple reactive RDW techniques promote broad adoption, while predictive or scripted approaches improve user experience under suitable environmental constraints or RL-based locomotion predictions.

Meta-strategies exist that combine various RDW methods. Azmandian et al. \cite{Azmandian22a} proposed Combinatorially Optimized Path-Planned Exploration Redirector (COPPER), which integrates offline static planning methods (scripted), online dynamic planning methods (predictive), and online reactive methods when static plans fail and predictions are unfeasible. COPPER outperformed S2C and FORCE in simulations~\cite{Azmandian22a}. Lee et al. \cite{Lee24d} presented Selective Redirection Controller (SRC), which dynamically switches among four state-of-the-art RDW controllers (S2C, P2R, SRL, and Williams et al. \cite{Williams21a}'s alignment controller) using RL to chose the best controller. This approach, which penalizes resets and rewards distance from walls, outperformed SRL. While RL-based methods can exceed APF methods in performance, they require more development time and resources due to training and tuning. Nevertheless, within each meta-strategy, each strategy can be considered its own RDW method and broken down into components via our framework. This is an interesting and promising avenue for future work, possibly including environment-based methods.

Concerns remain about RDW techniques and their evaluation, highlighting future research directions. Hirt et al. \cite{Hirt22} compared various reactive techniques with simulated and real user paths, finding that while simulation can be useful, many techniques did not improve significantly when tested against real user locomotion, with simulated behavior often exaggerated~\cite{Hirt22}. This suggests that techniques may perform better in situations with perfect reorientation (users reoriented with 100\% efficiency) and simple straight paths, but human locomotion, influenced by head motions during walking, is less predictable~\cite{Hirt22, Krueger23}. Thus, while simulation provides initial verification, real user testing is essential for understanding performance under non-ideal locomotion. In contrast, Azmandian et al. \cite{Azmandian22b} found minimal discrepancies between simulation and real-user performance, suggesting real-world deviations, such as slight head movement, are not significant. These contrasting findings show that both the simulations used and user behavior can influence the comparability of results. Most recent studies address this by incorporating live user tests, an approach that future research should continue to emphasize.

Bruder et al. showed RDW significantly affects cognitive load in both verbal and spatial tasks \cite{Bruder15}, indicating that high gains may impair task performance, while task cognitive demands can also alter users' motion. This factor limits RDW acceptance, particularly in smaller PEs where larger gains and more interruptions are needed~\cite{Williams07, Steinicke08a}. Although overt gain research shows that presence is not always broken by detectable gains, higher gains can still impact usability and cognitive load \cite{Bruder15, Schmitz18}. 
Langbehn et al. \cite{Langbehn18a} found that compared to other locomotion techniques, RDW provided superior spatial knowledge, was preferred over joystick movement, and had no significant impact on presence or task completion times.
Kim et al. reported that in dynamic environments RDW improved spatial knowledge, while in static ones, joystick control performed better. The differences mainly affecting object locations recall rather than object types identification or their correlations~\cite{Kim24}.
They also found longer task completion times for RDW compared to joystick and teleportation~\cite{Kim24}.
These findings suggest that while RDW can enhance user acceptance, particularly in scenarios where spatial knowledge is important, it may occasionally underperform in metrics like task completion time. 

In summary, RDW can enhance spatial knowledge and sense of presence compared to other locomotion techniques, but excessive gain levels may diminish presence and increase cognitive load. This indicates that RDW is a viable option when real walking is not possible in large environments, but its limitations should be carefully considered. 
For applications with high cognitive demands, such as verbal tasks, alternative methods like teleportation may be preferable. Future RDW research should evaluate a broader range of metrics including cognitive load, task performance, and presence—beyond mere redirection effectiveness. Enhancements such as gain masking techniques including distractors and non-visual stimuli should also be evaluated for impact on cognitive load against performance improvement as the introduction of new stimuli could further impact cognitive load on users. 

\begin{figure*}[t]
  \centering
  \includegraphics[width=\textwidth]{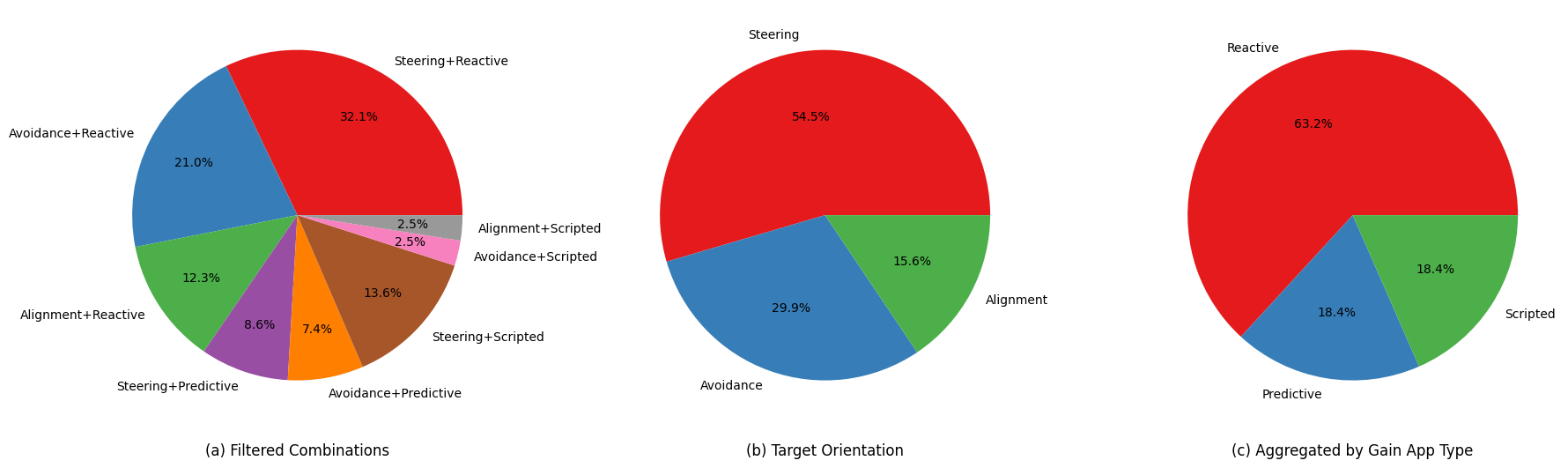}
  \caption{ (a) Percentage of works on RDW methods by combination of \textit{Gain Application} type and \textit{Target Orientation Calculation} method. (b) Percentage of works by  \textit{Target Orientation Calculation} type. (c) Percentage of works by \textit{Gain Application} type}
  \label{fig:precent_type_combo}
\end{figure*}

Figure~\ref{fig:precent_type_combo} (b) and (c) shows the percentage breakdown of research on RDW in our archive, categorized by \textit{Target Orientation Calculation} and \textit{Gain Application}, respectively. Majority of research focused on steering methods, with 54.5\% of papers in the archive discussing \textit{Heading Calculation} using steering based approaches. In contrast, the newer concept of alignment accounts for 15.6\% of the work, which is a notable share given its recency, indicating its potential and growing interest in methods that support passive haptics. Avoidance based methods, representing 29.9\% of the research, also demonstrate their usefulness, particularly as these approaches gained prominence around 2019 with the introduction of APFs for RDW~\cite{Bachmann19, Thomas19}. For \textit{gain application} type, reactive methods comprise the majority of the research, accounting for 63.2\% of the work, compared to 18.4\% each for predictive and scripted methods. 
This is likely due to reactive methods being the easiest to generalize across any VE and PE, while also being simplest to implement, as they only consider the user's current state.
Predictive methods likely have the lowest percentage of work due design and implementation challenges, as most require constraints on the VE, limiting their generalizability.
Generalizable predictive methods have been developed and are gaining traction, yet often depend on RL-based locomotion prediction, making them less accessible, more complex, and costlier to implement. 
Moreover, these RL methods require effective training to generalize beyond the training environments. Scripted methods also account for a small percentage due to inherent limitations: while they can theoretically achieve best path and performance in ideal scenarios, they demand prior knowledge of the VE and PE, and impose many constraints to ensure a solvable graph for static path planning.
Scripted methods likely show the lowest saturation since they require an online component when imposed constraints are violated, necessitating reactive or predictive solutions. Reactive methods, being the simplest starting point with no additional input beyond heading calculation and general enhancements, have a higher saturation of work. Methods can later be modified into predictive or scripted approaches. Figure~\ref{fig:precent_type_combo} shows the percentage distribution of all combinations of \textit{Heading Calculation} and \textit{Gain Application} from our literature archive, including only papers specifically addressing these RDW methods, and excluding those discussing general enhancements or converting any RDW approach into a predictive method, as in \cite{Jeon24, Zank17}.
Overall, reactive and steering methods dominate the research landscape, consistent with early influential RDW methods, such as those by Razzaque et al. \cite{Razzaque05} focusing on reactive steering. 
The large volume of work in this category reflects frequent RDW expansions through modifications to S2C and S2O.
Steering remains dominant across all gain application types, likely because avoidance and alignment are more recent concepts. 
Our archive currently lacks predictive alignment methods, highlighting a potential gap in RDW research.
Most scripted methods focus on steering, 13.6\% are scripted-steering compared to only 2.5\% each for scripted-avoidance and scripted-alignment. This is logical, as scripted methods involve static planning, where determining an ideal steering path is most applicable, leading to a higher prevalence of scripted-steering techniques. 
Predictive-avoidance methods show a saturation closer to predictive-steering than reactive-avoidance to reactive-steering. This likely reflects the prevalence of ML-based RDW methods, which often focus on avoidance due to the simplicity of using collision penalties as a reward function.
Our concept of \textit{Target Orientation Calculation} and component-wise framework highlights research gaps and offers a more complete foundation for developing and selecting perception-based RDW techniques (see supplementary material for ample usage of the taxonomy).

Enhancements like multi-user support have been widely explored, but aspects such as multi-player fairness remain under investigated with only limited studies available. Future research could explore fair resets in co-located applications or with different combinations of \textit{Target Orientation Calculation} types and \textit{Gain Application} types. 
Expanded motions for RDW were found promising for single-user applications, but are currently underexplored for multi-user applications.
Similarly, techniques like gain masking have mostly been demonstrated as proofs-of-concept with static gains and could benefit from further integration and evaluation within RDW methods. 
Distractors have proven effective in RDW, although their usefulness depends heavily on context, working best in interactive and gaming scenarios.
Prior findings suggest distractors can also be successfully incorporated into less interactive experiences, provided the VR application is specifically designed around them \cite{Yu18}.
Therefore, many of these enhancements may require customized RDW solutions tailored to individual VR experiences rather than generic approaches. Various RDW toolkits were developed to simplify adoption for VR applications and advance RDW research. These toolkits offer essential components for implementing RDW, including evaluation techniques, and simulation frameworks~\cite{Azmandian16, Li21a, Liu21}. Our taxonomy is intended to inspire improvements to existing toolkits or creating new ones supporting a modular design of RDW techniques.

\section{Limitations and Future Work}
To focus on an in-depth review and taxonomy of perception-based RDW techniques, we excluded environment manipulation methods from this survey's scope. 
Some previous surveys classified environment manipulation techniques without locomotion gains as scripted RDW methods~\cite{Nilsson18a}.
Excluding these techniques from our survey and taxonomy may partly explain the low representation of scripted techniques in our archive. 
This indicates that scripted techniques may be more prevalent in research than our findings suggest. However, as these techniques do not involve locomotion gains, our archive still accurately reflects the saturation of scripted techniques that directly manipulate user locomotion.

The keywords used to build our paper archive were chosen to optimize results. However, a different set of keywords could yield different outcomes and reveal works missed in this survey. Additionally, studies on reset techniques may not have used the term "redirected walking", potentially resulting in some of them being unintentionally excluded from the archive. Our survey endeavors to provide a comprehensive taxonomy of perception-based techniques for RDW. Nevertheless, we acknowledge a potential limitation due to focusing primarily on high-impact papers. This may have led to excluding emerging research that could gain future importance. This exclusion criterion was intended to ensure quality and influence of the included paper corpus, yet might have inadvertently overlooked some research work that could contribute to a more nuanced understanding of the RDW field. Future work could build-upon our taxonomy and expand the inclusion criteria to encompass emerging research even from papers published in lesser-known venues or with fewer citations to enrich the presented taxonomy.

Classification of techniques by types of gains used was omitted from tables and data due to unclear or unavailable information in many papers. Although gains are a key component of RDW and part of the taxonomy, classifying techniques by gains proved challenging and may be of limited value, as many techniques employ multiple gains types. Nevertheless, gains are a core component of RDW functionality, and as such need in future work to be expanded up on within the taxonomy and survey. 
Furthermore, Figure~\ref{fig:precent_type_combo} excludes works classified as general enhancements if their primary contribution was an enhancement applicable to any technique, regardless of heading calculation or gain application type. This applies even if the paper used a known heading calculation or gain application type, such as S2C, as the basis for proposing the enhancements.
If such works were classified based on the techniques they used, even when the enhancement is broadly applicable beyond specific heading calculation and gain application types, the distribution of work would likely appear different.

\section{Conclusion}
We propose a taxonomy for perception-based RDW techniques, emphasizing the key components essential for RDW algorithms. 
Our taxonomy introduces a newly proposed component for perception-based RDW: target orientation calculation. This addition affords a more comprehensive framework for perception-based RDW compared to previous frameworks that omitted this element.
The taxonomy resulted from a comprehensive literature review and classification of works published in principal research venues, offering a new perspective on RDW by introducing a new classification category based on target orientation calculation methods. 
We demonstrate how our taxonomy serves as a component-wise framework, enabling readers to design new perception-based RDW techniques or select suitable existing techniques for VR applications.
We explore key insights within each framework component and discuss general enhancements that can be broadly applied across various dissimilar methods.

We provide insights for future research within the context of our taxonomy, highlighting areas like predictive alignment methods, which were absent in the corpus. We hope our framework inspires the development of new RDW techniques, particularly through the explicit definition and identification of the target orientation calculation method, fostering the creation of new and improved approaches for calculating target orientation for perception-based RDW. 
We also hope that our taxonomy assists future researchers and developers in determining whether RDW suits their applications and identifying components that best fit their needs.

\printbibliography

\end{document}